\documentclass{article}

\usepackage[utf8]{inputenc}
\usepackage{amsmath}
\usepackage{fullpage}
\usepackage{relsize}
\usepackage{amssymb}
\usepackage{amsfonts}
\usepackage{bm}
\usepackage{mathabx}
\usepackage{graphicx}
\usepackage{tikz}
\usepackage{caption}
\usepackage{subcaption}
\usepackage{tikz-3dplot}
\usepackage{float}
\usepackage{xcolor}
\usepackage{pgfplots}
\usepackage{caption}
\usepackage{subcaption}
\usepackage{authblk}
\usepackage[english]{babel}
\usepackage{mathtools}
\usepackage{array}
\usepackage[normalem]{ulem}

\makeatletter
\newcommand{\thickhline}{%
    \noalign {\ifnum 0=`}\fi \hrule height 1.5pt
    \futurelet \reserved@a \@xhline
}
\newcolumntype{"}{@{\hskip\tabcolsep\vrule width 1.5pt\hskip\tabcolsep}}
\newcolumntype{?}{!{\vrule width 1.5pt}}
\makeatother

\usepackage[square,comma,numbers,sort&compress]{natbib}
\usepackage{hyperref}

\usetikzlibrary{arrows.meta}
\usetikzlibrary{calc}
\usetikzlibrary{shapes.misc}
\usepgfplotslibrary{fillbetween}

\definecolor{redmq}{RGB}{168,35,62}
\definecolor{purmq}{RGB}{129,37,92}
\definecolor{sanmq}{RGB}{214,210,196}
\definecolor{dredmq}{RGB}{118,35,47}
\definecolor{blumq}{RGB}{51,153,255}

\DeclareMathOperator{\erf}{erf}
\renewcommand{\Im}{\operatorname{Im}}
\DeclareMathOperator{\Arg}{Arg}
\DeclareMathOperator{\sech}{sech}
\renewcommand{\Re}{\operatorname{Re}}

\renewcommand{\i}{\mathrm{i}}
\renewcommand{\a}{\mathrm{e}}

\newcommand{\pdiff}[2]{\frac{\partial #1}{\partial #2}}

\newcommand\cjl[1]{{\color{magenta}{#1}}} 
\newcommand\change[1]{{\color{black}{#1}}} 

\hypersetup{colorlinks=true, linkcolor=redmq, citecolor=purmq, urlcolor=redmq, filecolor=redmq}
\pgfplotsset{compat=1.14}

\title{Nanoptera In Higher-Order Nonlinear Schr\"odinger Equations: Effects Of Discretization}
\author[1]{Aaron J. Moston-Duggan\footnote{Corresponding Author. Electronic address: aaron.moston-duggan@mq.edu.au}}
\author[2]{Mason A. Porter\footnote{Electronic address: mason@math.ucla.edu}}
\author[1]{Christopher J. Lustri\footnote{Electronic address: christopher.lustri@mq.edu.au}}
\affil[1]{Department of Mathematics and Statistics, 12 Wally's Walk, Macquarie University, New South Wales 2109, Australia}
\affil[2]{Department of Mathematics, University of California, Los Angeles, CA 90095, USA and Santa Fe Institute, Santa Fe, NM 87501, USA}
\date{}

\begin{document}
\maketitle



\begin{abstract}

We consider generalizations of nonlinear Schr\"odinger equations, which we call ``Karpman equations'', that include additional linear higher-order derivatives. Singularly-perturbed Karpman equations produce generalized solitary waves (GSWs) in the form of solitary waves with exponentially small oscillatory tails. Nanoptera are a special case of GSWs in which these oscillatory tails do not decay. Previous research on continuous third-order and fourth-order Karpman equations has shown that nanoptera occur in specific settings. We use exponential asymptotic techniques to identify traveling nanoptera in singularly-perturbed continuous Karpman equations. We then study the effect of discretization on nanoptera by applying a finite-difference discretization to continuous Karpman equations and studying traveling-wave solutions. The finite-difference discretization turns a continuous Karpman equation into an advance--delay equation, which we study using exponential asymptotic analysis. By comparing nanoptera in these discrete Karpman equations
with nanoptera in their continuous counterparts, we show that the oscillation amplitudes and periods in the nanoptera tails differ in the continuous and discretized equations. 
We also show that the parameter values at which there is a bifurcation between nanopteron and decaying oscillatory solutions depends on the choice of discretization.
Finally, by comparing different higher-order discretizations of the fourth-order Karpman equation, we show that the bifurcation value tends to a nonzero constant for large orders,
rather than to $0$ as in the associated continuous Karpman equation.

\end{abstract}

\section{Introduction}\label{S.Intro}

Nonlinear Schr\"{o}dinger (NLS) equations with higher-order dispersion terms, which we call ``higher-order nonlinear Schr\"{o}dinger equations'', arise in a variety of physical situations. In particular, recent experiments have demonstrated that higher-order dispersive effects are important in laser physics \cite{blanco2016pure,tam2019stationary,PhysRevA.101.043822,runge2020pure,PhysRevResearch.3.013166}. Experimental and theoretical studies of laser pulses and optical fibers have revealed that fourth-order dispersion can produce both bright \cite{PhysRevA.103.063514,posukhovskyi2020normalized,demirkaya2019numerical,akhmediev1994radiationless,piche1996bright,karlsson1994soliton,karpman1991influence1,karpman1991influence2,karpman1996lyapunov,karpman1994solitons,stefanov2022mixed} and dark \cite{alexander2021dark} solitary waves in photonic systems, whereas third-order dispersion prevents stable solitary-wave behavior \cite{zakharov1998optical}. Solitary waves have also been observed in experiments in systems that are modeled by sixth-order, eighth-order, and tenth-order dispersion, and they have been studied numerically for even dispersion orders of up to the sixteenth order \cite{PhysRevResearch.3.013166}. 

Motivated by early studies of higher-order NLS equations by V.~I. Karpman \cite{karpman1993radiation,karpman1993stationary,karpman1994solitons}, we refer to NLS equations with a linear higher-order derivative term as ``Karpman equations''. Karpman studied equations with third-order and fourth-order dispersion, and he showed that solutions to these higher-order NLS equations can possess ``generalized solitary waves" (GSWs), including nanoptera.  \textcolor{black}{Calvo and Akylas~\cite{calvo1997formation} studied a related singularly-perturbed third-order NLS equation that possesses both GSW and 
solitary-wave solutions.}

We use the term GSWs to describe stationary or traveling waves that have a central wave core and exponentially small oscillatory tails on one or both sides of the core~{\cite{Boyd1990ANC,boyd1999devil,boyd2012weakly}}. One type of GSW is a ``nanopteron'', in which the oscillatory tails do not decay spatially but instead persist indefinitely with a non-vanishing amplitude. Karpman \cite{karpman1994solitons} showed that stationary solutions of the fourth-order NLS equation can possess nanoptera. In the present paper, we use exponential asymptotic analysis to extend this result to non-stationary solutions. We then study the behavior of nanoptera and other GSWs in spatially and temporally discrete versions of Karpman equations. We refer to such systems as ``discrete Karpman equations''.

A key motivation for our paper is to investigate the effects of discretization on exponentially small oscillations in GSWs. We find that discrete Karpman equations have a bifurcation parameter that determines whether or not the solutions have nanoptera. We also find that the choice of discretization affects the value at which this bifurcation occurs, the amplitude of the oscillatory tails, and the period of those tails.


\subsection{Generalized Solitary Waves and Karpman Equations} \label{GSCH}

We study GSWs in Karpman equations. In studies of GSWs, it is common to distinguish between ``nanoptera'' and ``radiatively decaying GSWs'' \cite{boyd1999devil,boyd2012weakly}. Radiatively decaying GSWs have exponentially small oscillations that decay in space away from a wave core; therefore, they are exponentially localized in space. By contrast, nanoptera have exponentially small oscillations with an amplitude that does not decay in space, so nanoptera are not exponentially localized. Instead, they are localized up to algebraic powers of a parameter. We show schematic illustrations of these behaviors in Figure \hyperlink{fig1}{1}. See \cite{boyd1999devil,boyd2012weakly} for further details about GSWs. \textcolor{black}{As discussed in \cite{duncan1993solitons}, one can also think of a two-sided nanopteron as a solitary wave that is traveling on an {oscillatory} background.}

\hypertarget{fig1}{}
  \begin{figure}[tb] 
	\captionsetup[subfigure]{justification=centering}
        \centering
        \begin{subfigure}[b]{0.32\textwidth}
            \centering
        \includegraphics[width=0.95\textwidth]{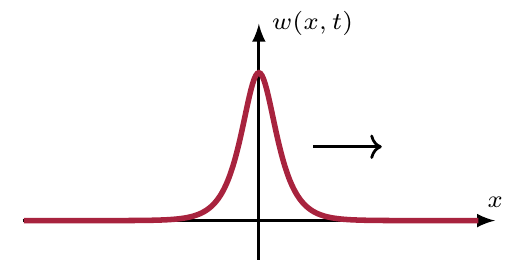} ~
        \caption{A solitary wave}
        \captionsetup{justification=centering}
        \end{subfigure}
        \begin{subfigure}[b]{0.32\textwidth}  
            \centering 
            \includegraphics[width=0.95\textwidth]{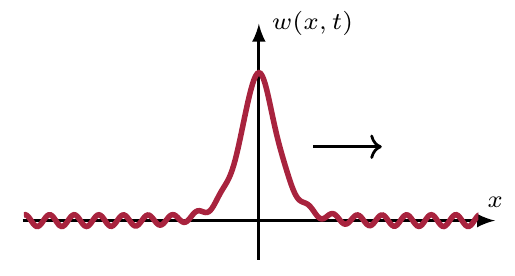}~
            \caption{A nanopteron}
            \captionsetup{justification=centering}
        \end{subfigure}
        \begin{subfigure}[b]{0.32\textwidth}   
            \centering 
            \includegraphics[width=0.95\textwidth]{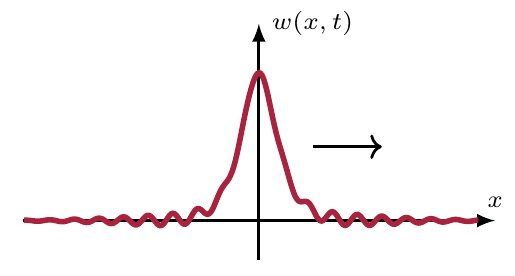}
            \caption{A radiatively decaying GSW}
            \captionsetup{justification=centering}
        \end{subfigure}
    \caption{Solitary waves and generalized solitary waves (GSWs), which we denote by $w(x,t)${.} (a) A solitary wave{, which} is spatially localized. 
    (b) A symmetric nanopteron, which has non-vanishing, exponentially small oscillations in its tails. (c) A radiatively decaying GSW, which has spatially localized oscillations that are exponentially small.
    }
   \label{fig1}
\end{figure}

GSWs typically occur in singularly-perturbed systems in which the associated unperturbed system has solitary-wave solutions \cite{boyd1999devil}; such systems can be differential equations or discrete (e.g., lattice) equations. 
For example, the cubic nonlinear Schr\"{o}dinger (NLS) differential equation
\begin{equation}\label{GNLS}
		\mathrm{i}\pdiff{\psi}{t} + \frac{1}{2}\pdiff{^2\psi}{x^2} +\left|\psi\right|^2\psi= 0
\end{equation}
has solitary-wave solutions, such as a focusing solitary wave (which we describe in Section \ref{legsw}). A generalization of the cubic NLS is the continuous fourth-order Karpman equation \cite{karpman1994solitons}
\begin{equation} \label{KARP}
	\mathrm{i}\pdiff{\psi}{t} + \frac{1}{2}\pdiff{^2\psi}{x^2}+\frac{\epsilon^2\lambda}{2}\pdiff{^4\psi}{x^4} +\left|\psi\right|^2\psi= 0\,. 
\end{equation}
This equation is a singularly-perturbed variant of the cubic NLS equation \eqref{GNLS} because of its linear fourth-order derivative term. Karpman expanded its solution in the limit $\epsilon\rightarrow0$ \cite{karpman1994solitons}, and he found that solitary-wave solutions exist for $\lambda<0$ and that nanopteron solutions exist for $\lambda>0$. In
\cite{karpman1994solitons}, he also {derived} an asymptotic description of the exponentially small oscillatory tails of the stationary nanopteron solutions.

GSWs have been studied both numerically \cite{karpman1991influence2,karpman1999evolution} and analytically \cite{karpman1991influence1,karpman1993radiation,karpman1994solitons,karpman1998evolution} in various continuous Karpman equations. These studies {derived} asymptotic descriptions of nanoptera in various {continuous} Karpman equations; in each case, it was determined that nanoptera exist for a countably infinite set of parameter values.

Discrete NLS equations also support solitary waves  \cite{dmitriev2007exact,pelinovsky2008stability,Ji2019,Zhang_2012,CUEVAS200967,Rothos_2008,PELINOVSKY20051,fitrakis2007,ZHU2020132326} and GSWs \cite{melvin2006radiationless,MELVIN2008551,melvin2009discrete,oxtoby2007moving,MA201793,alfimov2019standing,alfimov2019,PELINOVSKY200722,Syafwan_2012,kivshar1993}. See \cite{kevrekidis2009discrete,kevrekidis2001} for reviews of discrete NLS equations. Motivated by the widespread application of discrete NLS equations, we study lattice Karpman equations that are discrete both in space and in time. 
\textcolor{black}{In our investigation, we convert the lattice {equations into advance--delay equations and examine the behavior of GSWs in the resulting systems}. With this approach, we are able to explicitly study the effects of discretization on nanoptera.}

\textcolor{black}{We begin with a lattice equation
\begin{align}\nonumber
	\frac{\i}{2\tau}(w_{m,n+1} - w_{m,n-1}) &+ \frac{1}{2h^2}(w_{m+1,n} - 2w_{m,n} + w_{m-1,n}) + |w_{m,n}|^2w_{m,n}\\
&+ \frac{ \lambda\epsilon^2}{2h^4}(w_{m+2,t} - 4w_{m+1,t} + 6w_{m,n} - 4 w_{m-1,n} + w_{m-2,n}) = 0 \,, \quad m,n\in \mathbb{Z} \,, \label{lattice-karp}
\end{align}
where $\tau$, $h$, $\epsilon$, and $\lambda$ are parameters. We refer to equation \eqref{lattice-karp} as a ``lattice {fourth-order} Karpman equation''.} 

\textcolor{black}{Let $t = \tau n$, $x = h m$, and write ${w}_{m,n} = \psi(x,t)$, where $x$ and $t$ are continuous variables. These transformations yield the second-order central-difference fourth-order Karpman (CDK2) advance--delay equation
\begin{align}\label{TSIN}
	\i \frac{\psi(x,t+\tau)-\psi(x,t-\tau)}{2\tau} &+ \frac{1}{2}\frac{\psi(x+{h},t)-2\psi(x,t)+\psi(x-{h},t)}{{h}^2} + |\psi(x,t)|^2\psi(x,t) \nonumber \\  
	&+ \frac{\epsilon^2\lambda}{2}\frac{\psi(x+2{h},t) - 4\psi(x+{h},t) + 6\psi(x,t) - 4\psi(x-{h},t) + \psi(x-2{h},t)}{{h}^4}  = 0\,.	
\end{align}
One can also derive this equation by applying a central-difference discretization to the {continuous} fourth-order Karpman equation \eqref{KARP} with a temporal step size of $\tau$ and a spatial step size of $h$.} {We use the term ``discrete Karpman equations'' to refer to advance--delay equations that we obtain either by scaling a lattice equation or by applying a discretization to a continuous Karpman equation.}
 
Joshi and Lustri~\cite{joshi2019generalized} showed that discretization typically results in singularly-perturbed advance--delay equations and found that GSWs in discrete Korteweg--de Vries (KdV) equations behave differently from GSWs in the continuous KdV equation. This observation motivates the question of whether or not GSWs depend on the choice of discretization. In particular, we seek to determine whether or not changing the discretization affects which parameter values have associated nanoptera solutions. We will study advance--delay equations that arise from several different finite-difference discretizations (in addition to the one in \eqref{TSIN}), and we will compare the resulting GSWs.

The oscillatory tails of nanoptera are exponentially small in some asymptotic parameter, so they are inaccessible to standard asymptotic power-series methods. Researchers use exponential asymptotic techniques to describe exponentially small behavior. These techniques have been employed to examine the exponentially small oscillatory tails of GSWs and nanoptera both in continuous systems and in discrete systems that arise from lattice equations. Prior research includes studies of GSWs in the KdV equation \cite{Grimshaw2010,grimshaw1995fifth,trinh2010,joshi2019generalized,pomeau1988structural}, Toda chains  \cite{lustri2018nanoptera,vainchtein2016solitary,okada1990solitary,tabata1996stable}, Fermi--Pasta--Ulam--Tsingou (FPUT) chains \cite{faver2017nanopteronstegoton,faver2020micropterons,lustri2020nanoptera,HOFFMAN201733}, and woodpile chains \cite{deng2021nanoptera,deng2021nanoptera2,kim2015highly,Xu_2015}. Several of these studies have demonstrated that the oscillatory tails of GSWs arise from the ``Stokes phenomenon'', which describes the appearance and disappearance of exponentially small terms on the two sides of special curves that are known as ``Stokes curves'' \cite{alfimov2019standing,trinh2010,joshi2019generalized,lustri2018nanoptera,lustri2020nanoptera}.

\textcolor{black}{An important application of GSWs arises in the study of the so-called ``Peierls--Nabarro'' energy barrier \cite{aigner2003new,currie1980statistical,peyrard1984kink}, which was introduced to explain the behavior of moving kinks in lattices. Suppose that one models a kink as a traveling solitary wave. In certain lattice systems, such a kink exists if and only if it travels at least as fast as some minimum speed. If a traveling wave moves slower than this speed, then it is a GSW that generates radiation and eventually stop moving. Similarly, in the present paper, we demonstrate that our discrete Karpman equations have a parameter $\lambda$ that has an analogous effect on traveling-wave solutions. In the CDK2 equation \eqref{TSIN}, traveling waves with $\lambda \leq 1/4$ do not produce radiation in the form of persistent exponentially small waves, so they can propagate indefinitely. However, if $\lambda > 1/4$, a traveling wave is a nanopteron that creates exponentially small oscillations in its wake, so it radiates energy and cannot persist forever.}

In the present study of Karpman equations, we follow the exponential asymptotic procedure that was developed in \cite{chapman1998exponential,olde1995stokes} and adapted for advance--delay equations in \cite{joshi2015stokes,king2001asymptotics}. See \cite{dingle1973asymptotic,berry1988stokes} for discussions of the theoretical foundation and applicability of these exponential asymptotic methods.


\subsection{Exponential Asymptotic Analysis} \label{eaback}

We now outline how to use exponential asymptotic analysis to study GSWs in Karpman equations that include a small parameter $\epsilon$. The Karpman equations that we examine include linear higher-order derivatives that are singularly perturbed as $\epsilon \rightarrow 0$. We aim to determine the asymptotic behavior of the exponentially small oscillations in the GSWs. Classical asymptotic power series cannot describe exponentially small terms, such as the oscillatory tails of GSWs, because they are smaller than any power of $\epsilon$ in the limit $\epsilon \rightarrow 0$. If one instead uses exponential asymptotic methods to study such behavior, one finds that Stokes curves play an important role in the GSW dynamics.

Consider a singularly-perturbed Karpman equation, with a small parameter $\epsilon$, that has a traveling wave that satisfies the cubic NLS equation \eqref{GNLS} as its leading-order behavior in $\epsilon$. We denote this solution by $\psi_0$. Analytically continuing this traveling wave reveals that this solution is singular at a set of points in the complex plane. These singular points are the endpoints of Stokes curves in the solution of the Karpman equation \cite{stokes1864discontinuity}. Stokes curves divide the domain of a solution into distinct regions, which have different exponentially small asymptotic behavior. If a Stokes curve crosses the real axis, exponentially small terms must appear in the solution on one or both sides of the curve, so the traveling wave must be a GSW.

If we examine solutions of a Karpman equation on the two sides of a Stokes curve, we observe a rapid ``jump'' in the size of the exponentially small oscillatory tails of a GSW. This behavior is known as ``Stokes switching'', and one can use exponential asymptotic methods to explicitly compute the jump size of exponentially small terms across a Stokes curve. By understanding the behavior that arises from Stokes curves, one can calculate the asymptotic form of exponentially small oscillations in a GSW and determine which of these solutions are nanoptera. 

In a singularly-perturbed system whose solutions have exponentially small terms, asymptotic power series typically diverge \cite{dingle1973asymptotic}. One can use this observation to truncate the divergent series in a useful way. Choosing the truncation point to minimize the error of a power-series approximation typically results in an error that is exponentially small in the limit $\epsilon \rightarrow 0$ \cite{berry1988stokes,berry1989uniform,boyd1999devil}. This process is known as ``optimal truncation''. We denote the number of terms in an optimally truncated expansion by $N_{\mathrm{opt}}$. Applying optimal truncation to GSWs yields an expression that takes the form of a sum of an optimally truncated power series and an exponentially small remainder. The purpose of exponential asymptotics is to isolate the exponentially small remainder.

The exponential asymptotic approach that we use is similar to the {approach in \cite{joshi2015stokes, joshi2017stokes, joshi2019generalized,king2001asymptotics}}. See these studies for more general discussions of this approach. The present paper follows the approach that was first developed in \cite{chapman1998exponential,olde1995stokes}, and it uses the modifications for discrete systems that were developed in \cite{joshi2015stokes,king2001asymptotics}. 

To do exponential asymptotics, we first expand a solution as an asymptotic series 
\begin{equation}\label{eaasympser}
	\psi\sim\sum\limits_{j=0}^\infty\epsilon^{ j} \psi_j \quad \text{as} \quad \epsilon \rightarrow 0\,.	
\end{equation}
We obtain the leading-order term of the series by setting $\epsilon = 0$. In our case, the leading-order term satisfies the cubic NLS equation \eqref{GNLS}. To generate a GSW in a solution of a Karpman equation, the leading-order behavior must be a solitary wave. We select the focusing solitary wave \cite{kevrekidis2015defocusing,sulem2007nonlinear} of the cubic NLS equation as the leading-order solution (see \eqref{LESNLS} below). 

Applying the series \eqref{eaasympser} to a Karpman equation and matching orders of $\epsilon$ produces a recurrence relation for $\psi_j$. Solving this recurrence relation requires differentiating and integrating earlier terms in the series. For series that include terms with singularities, this process causes a particular type of divergence that is known as a ``factorial-over-power'' divergence \cite{dingle1973asymptotic}. 

In nonlinear problems, it is difficult to obtain an exact expression for the terms $\psi_j$. However, knowing the functional form of the divergent behavior enables one to obtain an asymptotic description of the so-called ``late-order'' terms $\psi_j$ as $j \rightarrow \infty$. One can capture this divergent behavior using an appropriate ansatz, such as in the form that was proposed by Chapman~\cite{chapman1996non} based on the work of Dingle~\cite{dingle1973asymptotic}. In this procedure, one assumes that the asymptotic behavior of the late-order terms as $j \to \infty$ is a sum of terms of the form
\begin{equation} \label{ealoteqn}
	\frac{\Psi\Gamma(\kappa j+\gamma)}{\chi^{\kappa j+\gamma}} \,,
\end{equation}
where $\kappa$ is the number of times that one needs to differentiate $\psi_{j-1}$ minus the number of times that one needs to integrate {$\psi_{j-1}$} to determine $\psi_j$, the scalar $\gamma$ is a constant, and $\Psi$ and $\chi$ are functions of the independent variables. Each term in the sum arises from a singularity of the leading-order solution $\psi_0$. In practice, one can determine the form of the factorial-over-power terms for each singularity independently, and one then sums the resulting contributions \cite{dingle1973asymptotic}.

The function $\Psi$ is known as the ``prefactor'' and the function $\chi$ is known as the ``singulant''. We require $\psi_j$ to be singular at singular points of the leading-order solution $\psi_0$. This gives the condition that $\chi = 0$ at points where $\psi_0$ is singular. As we increase $j$, the order of the singularity in $\psi_j$ increases. We find the functions $\Psi$ and $\chi$ by substituting \eqref{ealoteqn} into a Karpman equation and matching as $j\to\infty$. We select the value of $\gamma$ so that the late-order terms are consistent with the leading-order solution near the singularities.

We use the singulant of the late-order terms to locate Stokes curves in a solution. Dingle~\cite{dingle1973asymptotic} established that {if a series has a {power-series} expression such as \eqref{eaasympser}, then a Stokes curve that produces a jump in exponentially small behavior satisfies} 
\begin{equation} \label{STOKESCOND}
    \mathrm{Im}\{\chi\} = 0\,, \quad \mathrm{Re}\{\chi\} > 0 \,.
\end{equation}
The first condition is an equal-phase condition, and the second condition ensures that the exponential contribution is exponentially small (rather than exponentially large).

Using the late-order behavior, we optimally truncate an asymptotic series. To determine the optimal truncation point $N_{\mathrm{opt}}$, we apply a common heuristic that was described in \cite{boyd1999devil}. According to this heuristic, the optimal truncation point typically occurs at the series term with the smallest magnitude. We express the optimally truncated series as  
\begin{equation} \label{REMso}
	\psi = \sum_{j=0}^{N_{\mathrm{opt}}-1} \epsilon^{ j}\psi_j + \psi_{\text{exp}}\,, 
\end{equation}
where the remainder $\psi_{\text{exp}}$ is exponentially small in the limit $\epsilon\rightarrow0$. We obtain an expression for $\psi_{\text{exp}}$ by substituting \eqref{REMso} into the original Karpman equation. Away from the Stokes curve, we approximate $\psi_{\text{exp}}$ using the Liouville--Green (i.e., WKBJ) method \cite{bender2013advanced}. This approximation breaks down {around} the Stokes curve. {Around} the Stokes curve, we determine $\psi_\text{exp}$ using the form
\begin{equation} \label{earemint}
	\psi_\text{exp} \sim \mathcal{S}\Psi\a^{-\chi/\epsilon} \quad \text{as} \quad \epsilon \rightarrow 0\,,	
\end{equation}
where the Stokes multiplier $\mathcal{S}$ is a function of the independent variables. The Stokes multiplier changes rapidly in a neighborhood of width $\mathcal{O}(\sqrt{\epsilon})$ as $\epsilon\rightarrow 0$ around the Stokes curve. The Stokes multiplier tends to a different constant on the two sides of a Stokes curve, and it thereby captures the Stokes-switching behavior. 

From the form of the remainder \eqref{earemint}, we see that if $\mathrm{Re}\{\chi_x\} = 0$, then the exponential term $\psi_{\text{exp}}$ is purely oscillatory; it does not grow or decay. If the prefactor is constant or oscillates slowly with a constant amplitude, as is the case for Karpman equations, then the remainder has non-vanishing oscillations. Throughout the present study, we will apply this condition to classify exponentially small oscillations either as having a non-vanishing amplitude or as decaying in space. If the oscillations do not vanish, the solution is a nanopteron. 

We determine the exponentially small term $\psi_\text{exp}$ explicitly by substituting \eqref{earemint} into the expression for $\psi_\text{exp}$ and using matched asymptotic expansions \cite{olde1995stokes} to study the behavior of $\psi_{\text{exp}}$ in a region of width $\mathcal{O}(\sqrt{\epsilon})$ around the Stokes curve. 

\textcolor{black}{The method of \cite{chapman1998exponential,olde1995stokes} was developed originally to study ordinary differential equations. The techniques were subsequently adapted in \cite{king2001asymptotics} to study difference equations and differential--difference equations (such as Frenkel--Kontorova lattices); {this adapted method was used} in \cite{joshi2015stokes,joshi2017stokes} to study one-dimensional discrete equations. The method from \cite{chapman1998exponential,olde1995stokes} has also been extended to study partial differential equations \cite{body2005exponential,chapman2005exponential}, which can exhibit a rich variety of behavior (such as higher-order Stokes phenomena) \cite{howls2004higher}. In the present paper, we build on ideas from Joshi and Lustri \cite{joshi2019generalized}, in which this exponential asymptotic method was adapted to study solutions of lattice equations, which they represented as two-dimensional advance--delay equations.}

\change{Another exponential asymptotic method that has been used to derive asymptotic approximations for solutions of singularly-perturbed NLS equations (see \cite{calvo1997formation,yang2010nonlinear}) is based on {analyzing} solutions of the singularly-perturbed KdV equation in \cite{grimshaw1993note}. This method involves expressing the wave as the solution of an ordinary differential equation and then studying it using Fourier transforms. (Similar methods were also employed in \cite{boyd2012weakly}.) In the present paper, we use the method of \cite{chapman1998exponential,olde1995stokes} because it allows us to study solutions of our original partial differential equation, rather than an ordinary differential equation for a steady-state solution.}


\subsection{Outline of Our Paper}
 
Our paper proceeds as follows. In Section \ref{legsw}, we describe the focusing solitary-wave solution of the cubic NLS equation. We use this solution as the leading-order solution for all variants of the Karpman equations that we consider in the present study. In Section \ref{TKE}, we do an asymptotic analysis of the oscillatory tails in GSWs of the continuous fourth-order Karpman equation \eqref{KARP}. We also determine the conditions that permit GSWs to exist. We compare our asymptotic results both with numerical solutions and with the results of \cite{karpman1994solitons,karpman1999evolution}. In Section \ref{OLOS}, we generalize the results of Section \ref{TKE} to a family of continuous higher-order Karpman equations. In Section \ref{ch3}, we determine the asymptotic behavior of GSW oscillatory tails in the CDK2 equation \eqref{TSIN}. We also determine the conditions that permit GSWs to exist. In Section \ref{AFOHOD}, we extend the results of Section \ref{ch3} to discrete higher-order Karpman equations. We then compare these results to the asymptotic solutions of the continuous system in Section \ref{OLOS}. We determine conditions for the existence of GSWs. In Section \ref{HOCDKE}, we extend the results of Section \ref{ch3} to arbitrary-order central-difference discretizations of the continuous fourth-order Karpman equation \eqref{KARP}. In Section \ref{sec7}, we conclude and further discuss our results.

\section{Focusing Solitary-Wave Solution of the Cubic NLS Equation} \label{legsw}

In our study, we consider Karpman equations in which the highest derivative has a small parameter $\epsilon$ as a coefficient. In each of our examples, the leading-order solution $\psi_0$ satisfies the cubic NLS equation \eqref{GNLS}. 
We set the leading-order solution to be the focusing solitary wave
 \begin{equation} \label{LESNLS}
 	\psi_0 = A\sech\left(A\left(x-Vt\right)\right)\mathrm{e}^{\i\left(\left(A^2-V^2\right)t/2 +Vx\right)}\,,
\end{equation}
where $A$ is the amplitude and $V$ is the {speed} \cite{kevrekidis2015defocusing,sulem2007nonlinear}. 

To do exponential asymptotic analysis, we need to analytically continue the leading-order solution \eqref{LESNLS} into the complex plane; we then determine the locations of singularities. The solitary wave $\psi_0$ has singularities at 
\begin{equation} \label{CDSPTS}
 	x_{p} = Vt + {\frac{\i \pi (2p-1)}{2A}} \,, \quad  p\in\mathbb{Z}\,,
\end{equation}
where $x\in\mathbb{C}$ because of the analytic continuation. From \eqref{earemint}, we see that the exponentially small terms scale proportionally to $\a^{-\Re\{\chi\}/\epsilon}$ as $\epsilon\to 0$. 
 
The late-order terms in \eqref{ealoteqn} are dominated by the contributions with the smallest $|\chi|$. Therefore, for real values of $x$, the dominant late-order behavior must arise from the singularities that are closest to the real axis. Consequently, we only consider contributions from the singularities with $p=0$ and $p=1$. We 
denote these singularity locations by
\begin{equation} \label{singularities}    
	x_+ = Vt + {\frac{\i\pi}{2A}} \qquad \text{and} \qquad x_- = Vt - {\frac{\i\pi}{2A}}\,. 
\end{equation}
 
It is also helpful to obtain the asymptotic behavior of $\psi_0$ in the neighborhoods of the singular points. We use this information in our calculations of the late-order terms. Expanding $\psi_0$ about the singularities $x=x_{\pm}$ yields
\begin{equation} \label{ILOAV}
	\psi_0\sim  \mp\frac{ \i}{x-x_{\pm}}\mathrm{e}^{\i\left(\left(A^2-V^2\right)t/2 +Vx_{\pm}\right)} \quad \text{as} \quad  x\rightarrow x_{\pm}\,, 
\end{equation} 
where the upper and lower sign choices correspond.

\section{The Continuous Fourth-Order Karpman Equation} \label{TKE}


\subsection{Series Expansion} \label{FASE}

We now study the behavior of GSWs in solutions of the continuous fourth-order Karpman equation \eqref{KARP}. We expand $\psi$ as an asymptotic power series in $\epsilon^2$ and write
\begin{equation}
	\psi(x,t)\sim\sum_{j=0}^{\infty}\epsilon^{2j}\psi_j(x,t) \label{asyexp} \quad \text{as} \quad \epsilon\rightarrow 0\,.
\end{equation}
We substitute \eqref{asyexp} into \eqref{KARP} and match terms of size $\mathcal{O}(\epsilon^{2j})$ as $\epsilon\rightarrow 0$, this gives a recurrence relation for $\psi_j$. This relation is
\begin{equation} \label{epsmatch}
	\i\pdiff{\psi_{j}}{t} + \frac{1}{2}\pdiff{^2\psi_j}{x^2} + \frac{\lambda}{2}\pdiff{^4\psi_{j-1}}{x^4} + \frac{1}{2}\sum_{k=0}^j\psi_{j-k}\sum_{l=0}^{k}\left(\psi_l\overline{\psi}_{k-l}+\overline{\psi}_l\psi_{k-l}\right) = 0\quad \text{for} \quad j\geq1 \,, 
\end{equation}
where bars denote complex conjugation and we note that $|\psi|^2 = \psi \overline{\psi}$. 


\subsection{Late-Order Terms} \label{FOPD}

We follow \cite{chapman1998exponential} and propose an ansatz for the series terms in the limit $j\rightarrow\infty$ using the form in \eqref{ealoteqn}. This ansatz is a sum of terms of the form
\begin{equation} \label{LOT}
	\psi_j(x,t) \sim \frac{\Psi(x,t)\Gamma(2j+\gamma)}{\chi(x,t)^{2j+\gamma}}  \quad \text{ as } \quad j\rightarrow\infty\,. 
\end{equation}
We substitute the ansatz \eqref{LOT} into the recurrence relation \eqref{epsmatch} to obtain 
\begin{align}
	 &\frac{\Psi\Gamma(2j+\gamma+2)}{\chi^{2j+\gamma+2}}\left({\frac{\lambda}{2}\chi^4_x + \frac{1}{2}\chi^2_x}\right) \nonumber \\
	  & ~~~~~~ - \frac{\Gamma(2j+\gamma+1)}{\chi^{2j+\gamma+1}}\left({\i\Psi\chi_t+\Psi_x\chi_x+ \frac{1}{2}\Psi\chi_{xx}+2\lambda\Psi_x\chi_x^3+3\lambda\Psi\chi_x^3\chi_{xx}}\right) + \mathcal{O}(\psi_j) = 0 \quad \text{as} \quad j\rightarrow\infty \, , \label{LOTmatch}
\end{align}
where subscripts of $\chi$ and $\Psi$ denote partial differentiation.


\subsubsection{Singulant Equation}

  \begin{figure}[tb]
	\centering
	\includegraphics{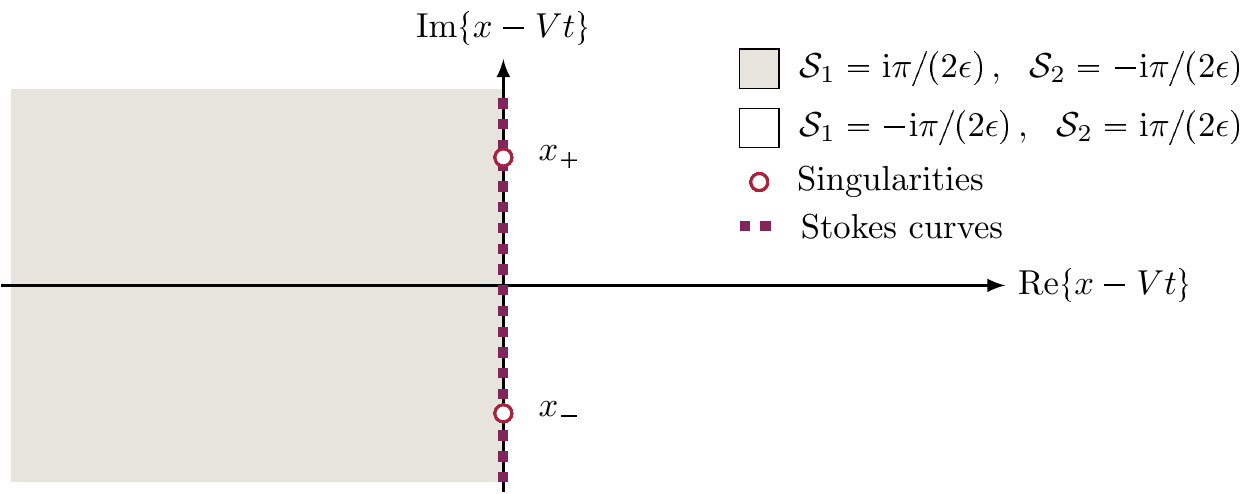}
	\caption{Stokes curves in the solution of the continuous fourth-order Karpman equation \eqref{KARP} with the leading-order focusing solitary wave solution  \eqref{LESNLS}. The Stokes curves both follow the dashed line (so they overlap). Each Stokes curve arises from one singularity of the leading-order solution. These singularities are located at $x = V t \pm \i\pi/(2A)$, which we show as red circles. The Stokes multipliers $\mathcal{S}_1$ and $\mathcal{S}_2$ take different values on the two sides of the Stokes curves. We shade one of these regions to distinguish between the two regions.
} 
	\label{stokmult}
\end{figure}

Matching at the leading order (i.e., matching terms of size $\mathcal{O}(\psi_{j+1})$ as $j\rightarrow\infty$) in \eqref{LOTmatch} yields the equation $\lambda\chi_x^4 + \chi_x^2 =0$. The singulant $\chi$ cannot be constant, so 
\begin{equation} \label{SING}
	\chi_x = \pm\i\sqrt{1/\lambda} \,.
\end{equation}
{The definition of the singulant requires that $\chi = 0$ at a singular point (see \eqref{CDSPTS}) of the leading-order solution.  We are interested in the singulants that are associated with the singularities $x = x_+$ and $x = x_-$ (see \eqref{singularities}). There is a singulant for each combination of sign choice and singularity location, so there are four singulants in total.}

It is helpful to distinguish between the solutions that are associated with positive and negative values of $\lambda$. To help our presentation, we define the notation $\alpha = \sqrt{1/\left|\lambda\right|}$, where
$\alpha \in \mathbb{R}^+$. If $\lambda>0$, the singulants are
\begin{gather} \label{POSLAM}
	\chi_{1} = \i\alpha(x-x_{+})\,, \quad
	\chi_{2} = -\i\alpha(x-x_{-})\,, \quad
	\chi_{3} = -\i\alpha(x-x_{+})\,,  \quad 
	\chi_{4} = \i\alpha(x-x_{-})\,, 
\end{gather}
where $\chi_x$ is imaginary for each singulant. If $\lambda<0$, the singulants are
\begin{gather} \label{NEGLAM}
	\chi_{1} = \alpha(x-x_{+})\,, \quad 
	\chi_{2} = \alpha(x-x_{-})\,, \quad 
	\chi_{3} = -\alpha(x-x_{+})\,,  \quad 
	\chi_{4} = -\alpha(x-x_{-})\,,  
\end{gather}
where $\chi_x$ is real for each singulant.

The form of the exponentially small oscillations is \eqref{earemint}. If $\Re\{\chi_x\}\neq0$ and $\Im\{\chi_x\}\neq0$, the singulants produce exponentially small oscillations that decay in space, so the solutions are radiatively decaying GSWs. If $\Re\{\chi_x\}=0$, the GSWs are nanoptera; they have exponentially small oscillations that do not decay in space, so they have non-vanishing amplitudes. Therefore, nanoptera exist only for $\lambda > 0$, for which $\mathrm{Re}\{\chi_x\} = 0$. 

Because we are particularly interested in nanoptera, we now restrict our attention to the case $\lambda > 0$. From the condition on $\mathrm{Re}(\chi)$ in \eqref{STOKESCOND}, we see that only $\chi_{1}$ and $\chi_2$ satisfy $\mathrm{Re}\{\chi\} > 0$ on the real axis. If $\lambda > 0$, these are thus the only singulants that generate the Stokes phenomenon. Therefore, they are the only singulants that generate exponentially small oscillations in a solution.

The singulant determines the locations of the Stokes curves (i.e., the ``Stokes structure'') of a problem. From \eqref{STOKESCOND}, we know that the Stokes curves satisfy $\Im\{\chi\}=0$, which occurs when $\Re\{x\}=Vt$. We show a schematic illustration of the Stokes curves in Figure \ref{stokmult}. This figure also shows the Stokes multipliers (which we compute in Appendix \ref{bapkar}), which have different values on the two sides of the Stokes curves. 


\subsubsection{Prefactor Equation}

Matching at the {next-to-leading} order (i.e., matching terms of size $\mathcal{O}(\psi_{j+1/2})$ as $j\rightarrow\infty$) in \eqref{LOTmatch} gives the prefactor equation. The singulant equation \eqref{SING} implies that ${\chi_{xx}}=0$, so the prefactor equation is 
\begin{equation}\label{PRE}
	\i\Psi\chi_t+ \Psi_x\chi_x+2\lambda\Psi_x\chi_x^3=0 \, . 
\end{equation}
Solving \eqref{PRE} yields
\begin{equation} \label{PREFF}
	\Psi(x,t) = f(t)\mathrm{e}^{-\i Vx} \,,
\end{equation}
where $f(t)$ is an arbitrary function that arises from integration with respect to $x$. We denote the prefactor and arbitrary function that are associated with {$\chi_{\nu}$ for $\nu \in \{1,2\}$ by $\Psi_{\nu}$ and $f_{\nu}${,} respectively.}

To determine $f(t)$, we match the late-order ansatz \eqref{LOT} with a local expansion of the solution near the singularity. We use the {behavior} in \eqref{ILOAV} as the leading order of the inner expansion, and we use that to generate the full local expansion. (See \eqref{INEX} in Appendix \ref{CTP}.) This analysis is similar to that in previous studies (such as \cite{chapman1998exponential,joshi2019generalized}) that used exponential asymptotic methods. {One cannot apply such a method 
directly to \eqref{KARP} because of the absolute-value term, which is not analytic.
 We overcome this issue by instead studying a
 system of two equations in which all terms are analytic functions.}

We rewrite the governing equation \eqref{KARP} as
  \begin{equation} \label{INNE1}
	\mathrm{i} \pdiff{\psi}{t} + \frac{1}{2}\pdiff{^2\psi}{x^2}+\frac{\epsilon^2\lambda}{2}\pdiff{^4\psi}{x^4} + \psi\overline{\psi}\psi = 0\,, 
\end{equation}
and we obtain a second equation by taking the complex conjugate of this {equation}. By {rescaling the two equations and replacing $\overline{\psi}$ with} a new variable, which equals the conjugate of $\psi$ on the real axis, we {obtain a system of equations in which all terms are analytic. We {then} analytically continue the solution of the resulting coupled system to {determine the behavior} of the analytic continuation of $\psi$ in the complex plane.} 
We present our detailed local analysis of the coupled system in Appendix \ref{CTP}.

Our local analysis gives
\begin{equation} \label{inpref1}
	f_1(t) = \alpha\Lambda\a^{\i\left(\left(A^2-V^2\right)t/2 + 2Vx_{+}\right)}
	\quad \text{ and } \quad  f_2(t) = \alpha\Lambda\a^{\i\left(\left(A^2-V^2\right)t/2 + 2Vx_{-}\right)}\,, 
\end{equation}
where $\Lambda$ is the limiting value of a recursive sequence. A computation gives $\Lambda \approx 4.494$.

The strength of the singularity in the late-order terms \eqref{LOT} must be consistent with the leading-order behavior \eqref{ILOAV} of the solitary-wave solution \eqref{LESNLS} as $x$ approaches the singularities 
$x=x_+$ and $x=x_-$. The singularities of the leading-order solution \eqref{LESNLS} are poles of order $1$. The leading-order behavior \eqref{ILOAV} is consistent with the late-order terms \eqref{LOT} if and only if $\gamma=1$. 

The complete form of the late-order terms is
\begin{equation} \label{fullLOT}
	\psi_j \sim  \frac{\alpha\Lambda\a^{\i\left(\left(A^2-V^2\right)t/2 -V(x- 2x_{+})\right)} \Gamma(2j+1)}{(\i\alpha(x-x_+))^{2j+1}} + \frac{\alpha\Lambda\a^{\i\left(\left(A^2-V^2\right)t/2 -V(x- 2x_{-})\right)} \Gamma(2j+1)}{(-\i\alpha(x-x_-))^{2j+1}}  \quad \text{ as } \quad  j\rightarrow\infty\,. 
\end{equation}


\subsection{Stokes Switching} \label{easec}

 \begin{figure}[tb]
	\begin{center}
	\includegraphics{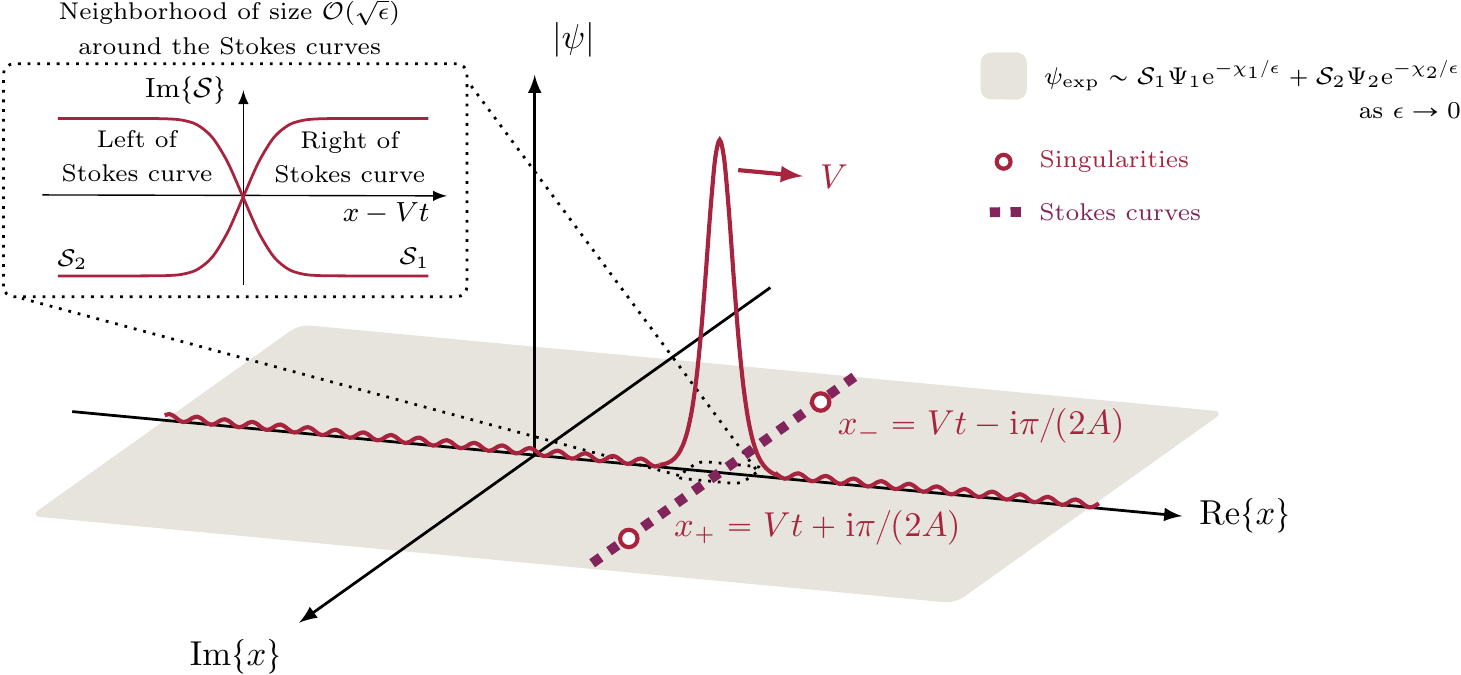}
 \end{center}
\caption{Stokes structure of a GSW solution $\psi$ of \eqref{KARP}. The singularities are endpoints of the Stokes curves. The Stokes curves are lines that are parallel to the imaginary axis; they move with velocity $V$ along with the core of {a GSW}. 
The Stokes multipliers $\mathcal{S}_1$ and $\mathcal{S}_2$ change their values in a neighborhood of width $\mathcal{O}(\sqrt{\epsilon})$ around the Stokes {curves}.
}
\label{stokket}
\end{figure}

We now follow the method that we described in Section \ref{eaback}; it is an application of the matched asymptotic-expansion approach of \cite{olde1995stokes}. We present complete details of our analysis in Appendix \ref{bapkar}. In that appendix, we show that the exponentially small contribution $\psi_{\mathrm{exp}}$ to the solution of \eqref{KARP} has non-vanishing oscillations of the form 
 \begin{equation} \label{EXPTER}
 	\psi_{\text{exp}} \sim \mathcal{S}_{1} \Psi_{1}\a^{-\chi_{1}/\epsilon} +  \mathcal{S}_{2} \Psi_{2}\a^{-\chi_{2}/\epsilon} \quad  \text{as} \quad \epsilon\rightarrow 0\,, 
 \end{equation}
 where {$\mathcal{S}_{\nu}$ and $\Psi_{\nu}$, respectively, are the Stokes multiplier and prefactor that are associated with $\chi_{\nu}$ for $\nu \in \{1,2\}$}. 
 We show a schematic illustration of this Stokes structure in Figure \ref{stokket}.

 Evaluating \eqref{EXPTER} yields
 \textcolor{black}{
 \begin{align} \nonumber
 	\psi_{\text{exp}} \sim -\frac{\i\pi\alpha\Lambda}{\epsilon} {\erf\left(\sqrt{\frac{\alpha\left|x - V t - \i\pi/{(2A)}\right|}{2\epsilon}}\Arg\left[\i\alpha\left(x - V t - \frac{\i\pi}{2A}\right)\right]\right)}& \\
 	\times\,  \a^{-\alpha\pi/(2A\epsilon)}\a^{\i\left(\left(A^2+3V^2\right)t/2 - Vx\right)}&\mu(x,t) \quad  \text{as} \quad \epsilon\rightarrow 0\,, \label{EXPTERSYM}
 \end{align}}
 where
 \begin{equation}
 	\mu(x,t) = \sinh\left(\frac{V\pi}{A}\right)\cos\left(\frac{\alpha}{\epsilon}(x-Vt)\right)  + \i\cosh\left(\frac{V\pi}{A}\right)\sin\left(\frac{\alpha}{\epsilon}(x-Vt)\right)
 \end{equation} 
 and ``Arg'' denotes the principle argument. The asymptotic behavior of the exponentially small oscillations for $x \in \mathbb{R}$ as $\epsilon\to 0$ is
\begin{equation}  \label{EXPTERSYMJ}
 	\psi_{\text{exp}}  \sim
 	\begin{dcases}
        \textcolor{white}{-}\frac{\i\pi\alpha\Lambda}{\epsilon} \a^{-\alpha\pi/(2A\epsilon)}\a^{\i\left(\left(A^2+3V^2\right)t/2-Vx\right)}\mu(x,t)\,,& x < V t\,, \\
        -\frac{\i\pi\alpha\Lambda}{\epsilon} \a^{-\alpha\pi/(2A\epsilon)}\a^{\i\left(\left(A^2+3V^2\right)t/2-Vx\right)}\mu(x,t)\,, & x > V t\,, \\
    \end{dcases}
\end{equation}
where the transition between the two asymptotic behaviors occurs in a narrow region of width $\mathcal{O}(\sqrt{\epsilon})$ {around}
the Stokes {curves}.


\subsection{Comparison of Results} 

\subsubsection{Numerical Comparison}\label{numer}

To validate the asymptotic description \eqref{EXPTERSYM} of a GSW, we compare its oscillation amplitude with the oscillation amplitude in numerical simulations of the continuous fourth-order Karpman equation \eqref{KARP}. 
We adapt a split-step method that was developed originally to solve the cubic NLS equation \cite{weideman1986split}. In each time step, split-step methods divide a computation into substeps for an equation's linear and nonlinear parts.
 To do this, we write the continuous fourth-order Karpman equation as
\begin{equation} \label{SSM1}
	\psi_t = \i\mathcal{L}(\psi) + \i\,\mathcal{N}(\psi)\,,  
\end{equation}
where the linear part $\mathcal{L}$ and the nonlinear part $\mathcal{N}$ are
\begin{equation} \label{SSM2}
	\mathcal{L}(\psi) = \frac{1}{2}\pdiff{^2\psi}{x^2} +  \frac{\lambda\epsilon^2}{2}\pdiff{^4\psi}{x^4}\quad \text{ and } \quad \mathcal{N}(\psi) = |\psi|^2\psi \,.
\end{equation}
The only difference from the approach in \cite{weideman1986split} is the presence of an extra fourth-order derivative term in the linear operator $\mathcal{L}$.

\begin{table}[tb] 
    \centering
    \begin{tabular}{|c|c|c|c|c|}
        \hline
        $\epsilon$ & Space domain & Spatial step size & Time domain & Temporal step size\\
        \hline $10^{-1}$   & $-1900\pi^2\epsilon \leq x \leq 1900\pi^2\epsilon$ & $\pi^2\epsilon/75$ & $0 \leq t \leq 1040\pi\epsilon$ & $\pi\epsilon/75$\\
        \hline $10^{-1.1}$ & $-1925\pi^2\epsilon \leq x \leq 1925\pi^2\epsilon$ & $\pi^2\epsilon/75$ & $0 \leq t \leq 980\pi\epsilon$ & $\pi\epsilon/75$ \\
        \hline $10^{-1.2}$ & $-1950\pi^2\epsilon \leq x \leq 1950\pi^2\epsilon$ & $\pi^2\epsilon/75$ & $0 \leq t \leq 920\pi\epsilon$ & $\pi\epsilon/75$ \\
        \hline $10^{-1.3}$ & $-1975\pi^2\epsilon \leq x \leq 1975\pi^2\epsilon$ & $\pi^2\epsilon/75$ & $0 \leq t \leq 860\pi\epsilon$ & $\pi\epsilon/75$ \\
        \hline $10^{-1.4}$ & $-2000\pi^2\epsilon \leq x \leq 2000\pi^2\epsilon$ & $\pi^2\epsilon/75$ & $0 \leq t \leq 800\pi\epsilon$ & $\pi\epsilon/75$ \\
        \hline
    \end{tabular}
    \caption{The parameter values that we use in our numerical computations of nanoptera.}
    \label{numparam}
\end{table}

 \begin{figure}[tb]
	\begin{center}
	\includegraphics[width = 0.75\linewidth]{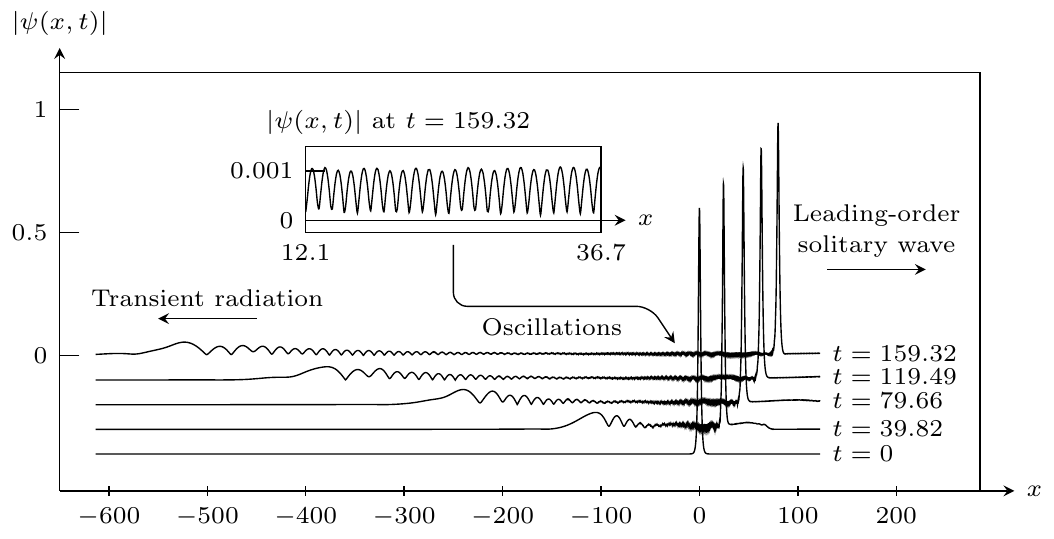} 
	\end{center}
	\caption{\textcolor{black}{Simulated solution of the continuous fourth-order Karpman equation \eqref{KARP} for $\epsilon = 0.1$ and an initial condition $\psi_0(x,0)$ that is given by \eqref{LESNLS}. We show the solution at $t = 0$ and at four evenly-spaced points in time {(which we round to two decimal places)}. 
		The vertical scale is for the solution at $t = 159.32$. We vertically offset the solutions at other times to clearly convey the wave propagation. The solution has a leading-order solitary wave that propagates to the right and transient radiation that propagates to the left. Exponentially small oscillations, which we magnify in the inset, form behind the leading-order solitary wave.}
	}
	\label{F.NumSim}
\end{figure}

Our initial condition $\psi_0(x,0)$ takes the form \eqref{LESNLS}. We evolve the solution in time on a spatially periodic domain. We adjust the domain parameters for each simulation (see Table \ref{numparam}). \textcolor{black}{We use a spatial domain that is large enough for finite-domain effects to not affect the nanopteron dynamics during our simulations. We show an example solution for $\epsilon = 0.1$ in Figure \ref{F.NumSim}. The leading-order solitary wave propagates to the right, and the solution's transient radiation propagates to the left. Exponentially small oscillations with {a} slowly varying amplitude occur between these features and trail the leading-order solitary wave. We run {our simulations for enough time to ensure that transient radiation} does not affect the behavior of the solution's oscillations.}

 \begin{figure}[tb]
	\begin{center}
	\includegraphics[width = 0.6\linewidth]{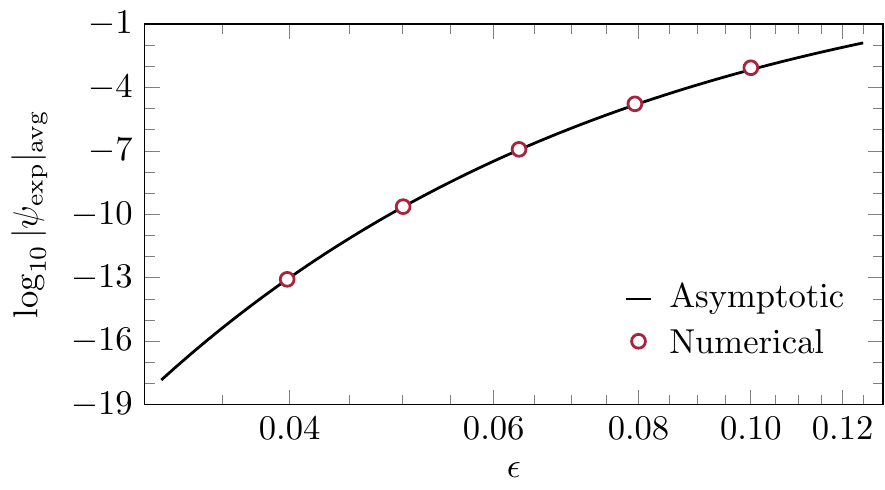} ~~~~~
	\end{center}
	\caption{The averaged amplitude $|\psi_{\text{exp}}|_{\text{avg}}$ of the exponentially small oscillatory tails of the asymptotic and numerical GSWs for $\lambda =1$, $A=1$, and $V=1$ using logarithmic axes. The asymptotic and numerical results are consistent, and the asymptotic estimate becomes more accurate as $\epsilon \rightarrow 0$.
	}
	\label{numerc}
\end{figure}

The oscillatory tail of the asymptotic solution \eqref{EXPTERSYM} has {a} {slowly varying periodic amplitude}
 \begin{equation} \label{EXPTERAMP}
 	\left|\psi_{\text{exp}}\right| \sim \frac{\pi\alpha\Lambda}{\epsilon} \a^{-\alpha\pi/{(2A\epsilon)}}{\sqrt{\sinh^2\left(\tfrac{V\pi}{A}\right)\cos^2\left(\tfrac{\alpha}{\epsilon}(x-Vt)\right)  + \cosh^2\left(\tfrac{V\pi}{A}\right)\sin^2\left(\tfrac{\alpha}{\epsilon}(x-Vt)\right)         }}
 	 \quad  \text{as} \quad \epsilon\rightarrow 0\,.
 \end{equation}
 To compare the theoretical oscillation amplitude from \eqref{EXPTERAMP} with our numerical results, we average the oscillation amplitude at the end of a simulation over two slowly-varying spatial periods. {We denote the averaged amplitude by $|\psi_{\text{exp}}|_{\text{avg}}$.} We show our asymptotic and numerical results in Figure \ref{numerc} for $\lambda =1$, $A=1$, and $V=1$ and a range of values of $\epsilon$. Our numerical simulations are consistent with our asymptotic estimates.


\subsubsection{Relation of our asymptotic solution
to the results in \cite{karpman1994solitons,karpman1999evolution}}

In Section \ref{easec}, we found that the {continuous} fourth-order Karpman equation \eqref{KARP} has nanopteron solutions
 only for $\lambda>0$. 
 Many years ago, Karpman \cite{karpman1994solitons} {derived} the same condition for the existence of stationary nanopteron solutions of the continuous fourth-order Karpman equation \eqref{KARP}. 

The amplitude \eqref{EXPTERAMP} of the oscillatory tails of traveling nanoptera varies slowly. 
{Each oscillatory tail of one of these nanoptera \eqref{EXPTERSYM} has} four periodic modes, with two each in space and time.
 The spatial modes have periods of $2\pi/V$ and $2\pi\epsilon/\alpha$, and the temporal modes have {periods} of $4\pi/(A^2+3V^2)$ and $2\pi\epsilon/(V\alpha)$, where $V \neq 0$ is the nanopteron speed. 
Previous research on stationary nanoptera (i.e., with $V = 0$) in the continuous fourth-order Karpman equation \eqref{KARP} \cite{karpman1994solitons} found that the amplitude of the oscillations is constant, with only one oscillatory mode in each of space and time, with a spatial period of $2\pi\epsilon/\alpha$ and a temporal period of $4\pi/A^2$. This is consistent with the asymptotic result \eqref{EXPTERSYM}; we recover the result from \cite{karpman1994solitons} by taking the limit $V \to 0$. One temporal mode and one spatial mode disappear in this limit because the associated periods become infinitely large.

A similar Karpman equation was studied numerically with $V \neq 0$ by Karpman and Shagalov \cite{karpman1999evolution}, who observed exponentially small oscillations with periodic amplitudes. This is consistent with the behavior of the asymptotic solution \eqref{EXPTERSYM}, which has the slowly varying periodic amplitude
\eqref{EXPTERAMP}.

\section{{Continuous} Higher-Order Karpman Equations} \label{OLOS}

\begin{figure}
	\captionsetup[subfigure]{justification=centering}
    \centering
      \begin{subfigure}[b]{0.3\textwidth}
            \centering
			\includegraphics{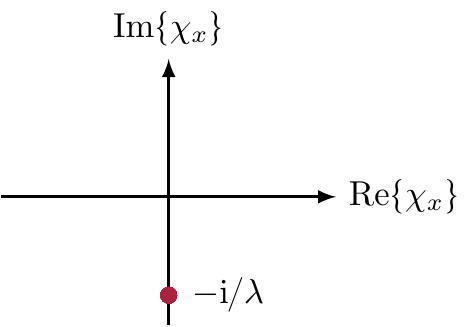}
			\caption{$k=1$~~~~~~~~~~~~~}
        \end{subfigure}
        \begin{subfigure}[b]{0.3\textwidth}  
                   \centering
                   \includegraphics{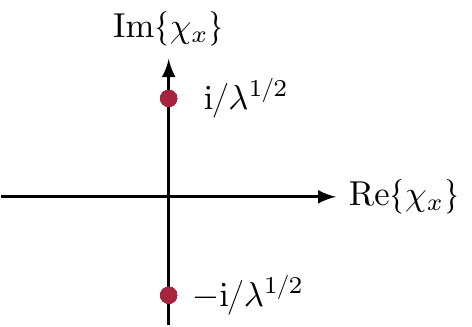}
			\caption{$k=2$~~~~~~~~~~~~~}
        \end{subfigure}
        \begin{subfigure}[b]{0.3\textwidth}   
                       \centering
                       \includegraphics{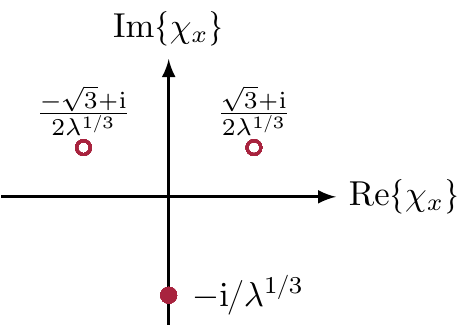}
	\caption{$k=3$~~~~~~~~~~~~~}
        \end{subfigure}       
        \begin{subfigure}[b]{0.3\textwidth}
            \centering
            \includegraphics{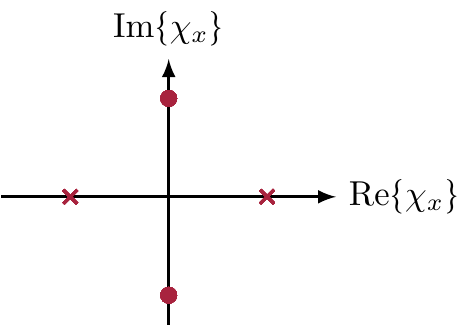}
	\caption{$k=4$~~~~~~~~~~~~~}
        \end{subfigure}
        \begin{subfigure}[b]{0.3\textwidth}  
            \centering 
            \includegraphics{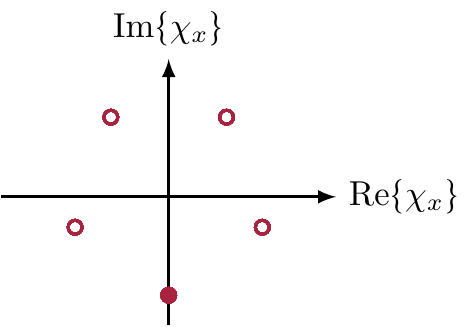}
	\caption{$k=5$~~~~~~~~~~~~~}
        \end{subfigure}
        \begin{subfigure}[b]{0.3\textwidth}   
            \centering 
            \includegraphics{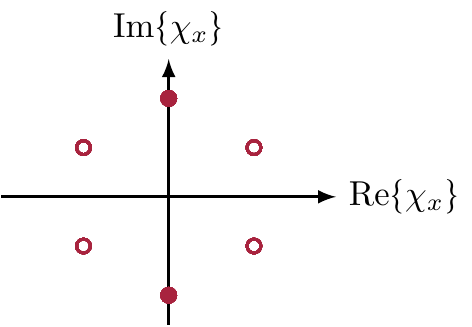}
	\caption{$k=6$~~~~~~~~~~~~~}
        \end{subfigure}
 \caption{Solutions of the singulant equation \eqref{HOS} for $\chi_x$ with $\lambda>0$ for several values of the parameter $k$. We show imaginary values of $\chi_x$ as filled circles; these values correspond to non-vanishing exponentially small oscillations in the solution of \eqref{CDNLSGHO}. Crosses and open circles correspond to real and complex values of $\chi_x$, respectively. These contributions lead to exponential decay in space. 
 }
   \label{HOSST}
\end{figure}

\begin{figure}
	\captionsetup[subfigure]{justification=centering}
        \centering
      \begin{subfigure}[b]{0.3\textwidth}
            \centering
            \includegraphics{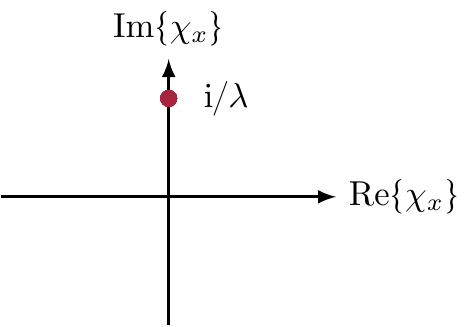}
	\caption{$k=1$~~~~~~~~~~~~~}
        \end{subfigure}
        \begin{subfigure}[b]{0.3\textwidth}  
            \centering
            \includegraphics{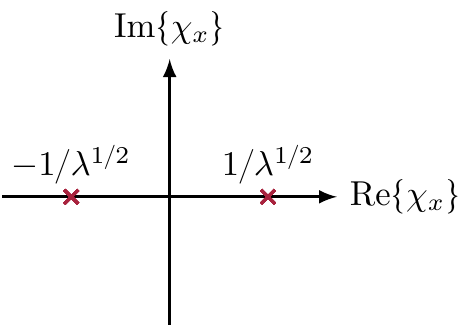}
	\caption{$k=2$~~~~~~~~~~~~~}
        \end{subfigure}
        \begin{subfigure}[b]{0.3\textwidth}   
                       \centering
                       \includegraphics{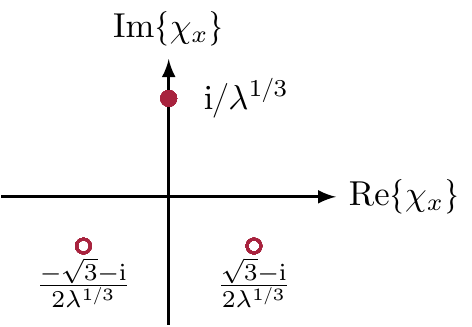}
	\caption{$k=3$~~~~~~~~~~~~~}
        \end{subfigure}       
        \begin{subfigure}[b]{0.3\textwidth}
            \centering
            \includegraphics{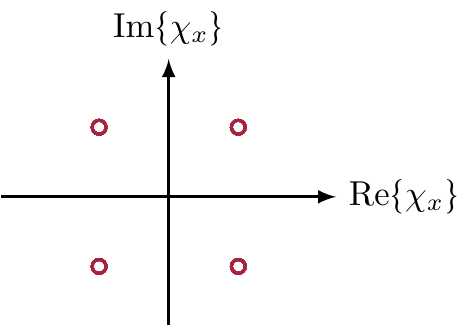}
	\caption{$k=4$~~~~~~~~~~~~~}
        \end{subfigure}
        \begin{subfigure}[b]{0.3\textwidth}  
            \centering 
            \includegraphics{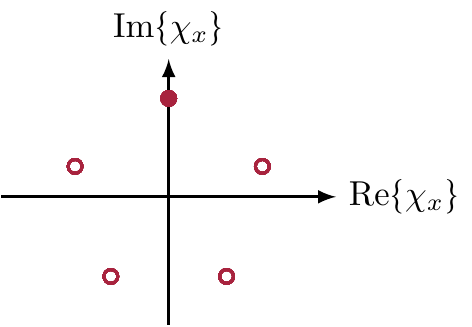}
	\caption{$k=5$~~~~~~~~~~~~~}
        \end{subfigure}
        \begin{subfigure}[b]{0.3\textwidth}   
            \centering 
            \includegraphics{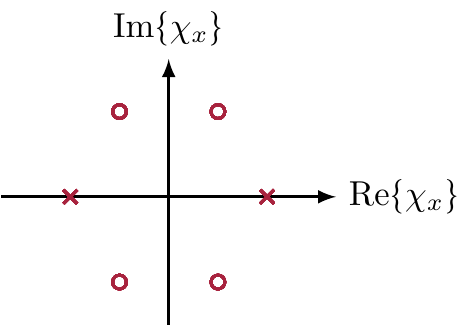}
	\caption{$k=6$~~~~~~~~~~~~~}
        \end{subfigure}
 \caption{Solutions of the singulant equation \eqref{HOS} for $\chi_x$ with $\lambda<0$ for several values of the parameter $k$. We show imaginary values of $\chi_x$ as filled circles; these values correspond to non-vanishing exponentially small oscillations in the solution of \eqref{CDNLSGHO}. 
 Crosses and open circles correspond to real and complex values for $\chi_x$, respectively. These contributions lead to exponential decay in space.}
   \label{HOSST2}
\end{figure}

\begin{figure}[tb]
	\centering
	\includegraphics{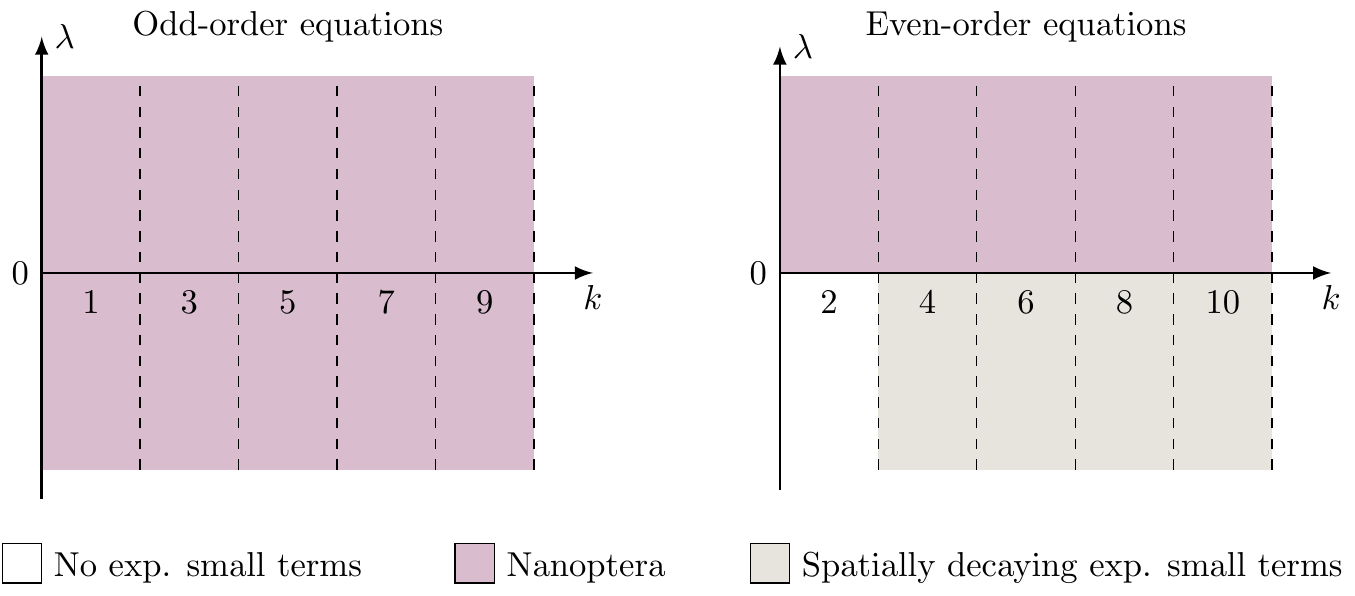}
	\caption{Bifurcation diagram of the traveling-wave solutions of the {continuous} higher-order Karpman equation \eqref{CDNLSGHO} for different values of $k$ and $\lambda$, where the order of the differential equation is $k+2$. If $k$ is odd, we obtain nanoptera for all $\lambda$. If $k$ is even, {we obtain nanoptera only} for $\lambda > 0$. 
	The solution behavior changes at the bifurcation value $\lambda = 0$.
	}
	\label{gswbehav}
\end{figure}

We consider a family of {continuous} Karpman equations that we index by a parameter $k$. Each example of this equation, which we call the ``continuous higher-order Karpman equation'', has a singularly-perturbed linear term with a higher-order derivative. This equation is
\begin{equation}\label{CDNLSGHO}
	\i\frac{\partial\psi}{\partial t} + \frac{1}{2}\frac{\partial^2\psi}{\partial x^2} - \frac{\lambda\epsilon^k\i^k}{2}\frac{\partial^{k+2}\psi}{\partial x^{k+2}} + |\psi|^2\psi = 0 \quad \text{  as  } \quad  \epsilon\rightarrow0\,. 
\end{equation}
This equation reduces to {the continuous fourth-order Karpman equation} \eqref{KARP} when $k=2$. Solutions of equation \eqref{CDNLSGHO} with $k=1$ were examined in \cite{calvo1997formation,karpman1993stationary,karpman1993radiation,wai1990radiations,wai1986nonlinear}, with each study finding that the solutions have exponentially small oscillations. For example, Karpman~\cite{karpman1993radiation} found that nanoptera occur for all $\lambda\neq0$.

To determine which \eqref{CDNLSGHO} can support GSWs, we compute the singulant {for all positive integers $k$.}
Contributions {for} which $\chi_x$ is imaginary have solutions with exponentially small non-vanishing oscillations. Therefore, the associated GSWs are nanoptera.

We follow a similar method as in Section \ref{FASE} and expand the solution of \eqref{CDNLSGHO} as an asymptotic power series in $\epsilon^k$. The leading-order solution $\psi_0$ satisfies the cubic NLS equation, and we select the focusing solitary wave \eqref{LESNLS} as the leading-order behavior. Matching terms of size $\mathcal{O}(\epsilon^{kj})$ as $\epsilon \rightarrow 0$ produces a recurrence relation for the terms in the asymptotic series. As in Section \ref{FOPD}, we determine the late-order terms using the ansatz 
\begin{equation} \label{LOTHO}
	\psi_j(x,t) \sim \frac{\Psi(x,t)\Gamma(kj+\gamma)}{\chi(x,t)^{kj+\gamma}}  \quad \text{ as } \quad j\rightarrow\infty\,. 
\end{equation}

The singulant equation has $k$ solutions, which satisfy
\begin{equation}  \label{HOS}
	\chi_x^k = 
\begin{cases}
	~~\i^k/\lambda\,, \quad\hspace{0.02cm} k \text{ is even} \\
	-\i^k/\lambda\,, \quad k \text{ is odd}\,.
\end{cases}
\end{equation}
We show values of $\chi_x$ from \eqref{HOS} for $\lambda>0$ and $\lambda<0$ in Figures \ref{HOSST} and \ref{HOSST2}, respectively. 

For even values of $k$ and $\lambda>0$, there are always {imaginary values of $\chi_x$ that satisfy \eqref{HOS}; these values are $\chi_x = \pm\i/\lambda^{1/k}$}. We show this for $k = 2$, $k = 4$, and $k = 6$ in Figure \ref{HOSST}. Any traveling-wave solution of \eqref{CDNLSGHO} for even values of $k$ with $\lambda > 0$ must be a nanopteron. For $k\geq4$, there are also complex values of $\chi_x$ with nonzero real parts that satisfy \eqref{HOS}.
 The imaginary values of $\chi_x$ yield non-decaying oscillations, and the other complex values of $\chi_x$ yield radiatively decaying oscillations. 
The solutions are still nanoptera because of the presence of the non-decaying oscillations.

For even values of $k$ and $\lambda<0$, there are no purely imaginary values of $\chi_x$ that satisfy \eqref{HOS}; all values of $\chi_x$ are real or complex with nonzero real parts. We show this for $k = 2$, $k = 4$, and $k = 6$ in Figure \ref{HOSST2}. These parameter values do not permit nanopteron solutions of \eqref{CDNLSGHO}. Instead, all exponentially small terms in {traveling-wave solutions of \eqref{CDNLSGHO}} decay exponentially in space away from the central wave core. 

For odd values of $k$ and any $\lambda$, there is one imaginary value of $\chi_x$ that satisfies \eqref{HOS}. We show this for $k = 1$, $k = 3$, and $k = 5$ in Figures \ref{HOSST} and \ref{HOSST2}. If $\lambda > 0$, then $\chi_x = \i/\lambda^{1/k}$ satisfies \eqref{HOS}; if $\lambda < 0$, then $\chi_x = -\i/\lambda^{1/k}$ satisfies \eqref{HOS}. Therefore, for odd $k$, any traveling-wave solution is a nanopteron. If $k\geq 3$, there are also complex values of $\chi_x$ with nonzero real parts that satisfy \eqref{HOS}.
 These solutions are associated with oscillations that decay radiatively in space away from the wave core.

We summarize the wave behavior of the continuous higher-order Karpman equation \eqref{CDNLSGHO} in Figure \ref{gswbehav}. For $k=1$, these results are consistent with previous research \cite{karpman1993stationary,karpman1993radiation,wai1990radiations,wai1986nonlinear} that found nanopteron solutions when $k=1$. It also confirms the observation in \cite{karpman1993radiation} that there are no bifurcations in $\lambda$ when $k=1$.

\section{{The} Second-Order Central-Difference Fourth-Order Karpman Equation {(CDK2)}} \label{ch3}

Joshi and Lustri~\cite{joshi2019generalized} studied the behavior of a discrete KdV equation and found that discretization can alter the parameter values in the system that permit nanopteron solutions. In this section, we study this phenomenon in advance--delay equations that arise from lattice Karpman equations. We determine whether or not we can recover the nanopteron solutions of continuous Karpman equations in a {discrete} system. We first study {the discrete fourth-order Karpman equation \eqref{TSIN} 
that we obtain by discretizing the {continuous} fourth-order Karpman equation \eqref{KARP} using}
second-order central differences in time and space. We denote a time step by $\tau$ and a space step by $h$. We use ``CDK2'' to refer to {the discrete Karpman equation \eqref{TSIN}.} 

To retain both time and space terms in a {{discrete}} Karpman equation, we require the temporal and spatial increments of the discretization to have the same order as $\tau\to 0$ and $h \to 0$. The CDK2 equation \eqref{TSIN} has a small parameter $\epsilon$, so it is natural to set $h = \epsilon$. To ensure that the temporal and spatial increments have the same order, we introduce a scaling parameter $\sigma$ so that $\tau = \sigma \epsilon$. In the limit $\epsilon\rightarrow 0$, we apply a Taylor approximation about $\epsilon = 0$ to obtain the infinite-order singularly-perturbed partial differential equation
\begin{equation} \label{CDNLS4}
	\i \sum_{r=0}^{\infty} \frac{(\sigma \epsilon)^{2r}}{(2r+1)!}\frac{\partial^{2r+1}}{\partial t^{2r+1}}\psi + \left(1-4\lambda\right)\sum_{r=0}^{\infty} \frac{\epsilon^{2r}}{(2r+2)!} 
\frac{\partial^{2r+2}}{\partial x^{2r+2}}\psi + \lambda \sum_{r=0}^{\infty} \frac{2^{2r+2}\epsilon^{2r}}{(2r+2)!} 
\frac{\partial^{2r+2}}{\partial x^{2r+2}}\psi  + |\psi|^2\psi  = 0\,.	
\end{equation}


\subsection{Series Expansion} \label{cdfase}

We apply the asymptotic series expansion of $\psi$ from \eqref{asyexp} to the CDK2 equation \eqref{TSIN} and match terms of size $\mathcal{O}\left(\epsilon^0\right)$ as $\epsilon\rightarrow0$. As before, $\psi_0$ satisfies the cubic NLS equation, and we again select the focusing solitary wave \eqref{LESNLS} as the leading-order behavior.

Matching terms of size $\mathcal{O}(\epsilon^{2j})$ as $\epsilon\rightarrow 0$ in \eqref{CDNLS4} produces a recurrence relation for $j \geq 1$. It is
 \begin{align} \label{CDREC}
 	\i\sum_{r=0}^{j}	\frac{\sigma^{2r}}{(2r+1)!}\frac{\partial^{2r+1}}{\partial t^{2r+1}}\psi_{j-r}& + \left(1-4\lambda\right)\sum_{r=0}^j \frac{1}{(2r+2)!}\frac{\partial^{2r+2}}{\partial x^{2r+2}}\psi_{j-r} \nonumber \\
 	& + \lambda\sum_{r=0}^j\frac{2^{2r+2}}{(2r+2)!}\frac{\partial^{2r+2}}{\partial x^{2r+2}}\psi_{j-r} + \frac{1}{2}\sum_{r=0}^j\psi_{j-r}\sum_{l=0}^{r}\left(\psi_l\overline{\psi}_{r-l}+\overline{\psi}_l\psi_{r-l}\right) = 0\,. 
 \end{align}


\subsection{Late-Order Terms} \label{cdLOT}

To determine the asymptotic behavior of the series terms, we use the late-order ansatz \eqref{LOT} that we used in Section \ref{FOPD}. Applying the ansatz to \eqref{CDREC} and retaining the two largest orders for $j\rightarrow \infty$ yields 
\begin{align}\label{LOTCD}
	& \frac{\Psi\Gamma(2j+\gamma+2)}{\chi^{2j+\gamma+2}}\left(\left(1-4\lambda\right)\sum_{r=0}^j\frac{\chi_x^{2r+2}}{(2r+2)!} + \lambda\sum_{r=0}^j\frac{\left(2\chi_x\right)^{2r+2}}{(2r+2)!}\right)        \nonumber \\  & -\frac{\Psi\Gamma(2j+\gamma+1)}{\chi^{2j+\gamma+1}}\left(\frac{\i}{\sigma}\sum_{r=0}^j\frac{\left(\sigma \chi_t \right)^{2r+1}}{(2r+1)!} + \frac{\Psi_x}{\Psi}\left(\left(1-4\lambda\right)\sum_{r=0}^j \frac{\chi_x^{2r+1}}{(2r+1)!} + 2\lambda \sum_{r=0}^j \frac{(2\chi_x)^{2r+1}}{(2r+1)!}\right)\right) + \mathcal{O}(\psi_j) = 0. 
\end{align}


\subsubsection{Singulant Equation}

Matching at the leading order (i.e., matching terms of size $\mathcal{O}(\psi_{j+1})$ as $j\rightarrow\infty$) in \eqref{LOTCD} gives the following equation for the singulant:
\begin{equation}\label{SING2}
	\left(1-4\lambda\right)\sum_{r=0}^j\frac{\chi_x^{2r+2}}{(2r+2)!} + \lambda\sum_{r=0}^j\frac{\left(2\chi_x\right)^{2r+2}}{(2r+2)!} = 0\,. 
\end{equation}
The expression \eqref{SING2} holds in the asymptotic limit $j \rightarrow \infty$ because we are considering the behavior of late-order terms. Extending the upper bound of the summation to infinity introduces an exponentially small error into the singulant \cite{king2001asymptotics}. From the form of the remainder term \eqref{earemint}, we see that this introduces an error into $\psi_{\text{exp}}$ that is exponentially small in comparison to the already exponentially small oscillations. The final expression for the exponentially small oscillations has asymptotic errors that are larger than this in the limit $\epsilon \rightarrow 0$, so extending the summation range is valid asymptotically. 

Extending the range of summation in \eqref{SING2} yields the singulant equation
\begin{equation}\label{cdsing}
	2\lambda\cosh^2(\chi_x) + \left(1-4\lambda\right)\cosh(\chi_x) + 2\lambda-1 = 0 \, . 
\end{equation}
Solving \eqref{cdsing} gives 
\begin{equation}\label{2SINGS}
	\chi_x = 2\pi\i q \quad \text{and} \quad \chi_{x} = 2\pi\i q \pm \cosh^{-1}{\left(\frac{2\lambda-1}{2\lambda}\right)}\,, \quad \text{with} \quad q\in\mathbb{Z} \,.  
\end{equation}

Recall from Section \ref{FOPD} that nanoptera exist only if $\chi_x$ is imaginary. The singulants with $\chi_x = 2\pi\i q $ satisfy this condition; however, we will show in Section \ref{ch5pref} that the prefactors that are associated with these singulants must vanish. Consequently, these terms do not contribute to the asymptotic expression of the solution of the CDK2 equation \eqref{TSIN}.

The second set of values of $\chi_x$ from \eqref{2SINGS} are imaginary if $\lambda\geq 1/4$; otherwise, they have nonzero real parts. Therefore, the CDK2 equation has nanopteron solutions when $\lambda \geq 1/4$. This situation differs from that in the continuous fourth-order Karpman equation, which has nanopteron solutions when $\lambda > 0$. The discretization changes the value of $\lambda$ at which traveling waves in the solution become nanoptera. We are interested primarily in nanoptera, so we focus on the case $\lambda \geq 1/4$.

There are an infinite number of possible values for $\chi_x$; we index them by $q$. The dominant asymptotic behavior of the late-order terms is determined by the value of $q$ that maximizes the late-order terms in \eqref{LOT}. For real values of $x$, this value is $q = 0$. 
Therefore, for the rest of our analysis, we focus on the two contributions from the second set of solutions in \eqref{2SINGS} with $q = 0$.

To facilitate our presentation, we define $\beta = \Im\{\cosh^{-1}((2\lambda-1)/\lambda)\}$; note that $0<\beta\leq\pi$ for $\lambda \geq 1/4$.
 The two relevant singulants are 
\begin{equation} \label{cdSINSIM}
	\chi_1 = \i\beta(x-x_+) \quad \text{and} \quad \chi_2 = -\i\beta(x-x_-)\,.  
\end{equation}

\begin{figure}[tb]
	\centering
	\includegraphics{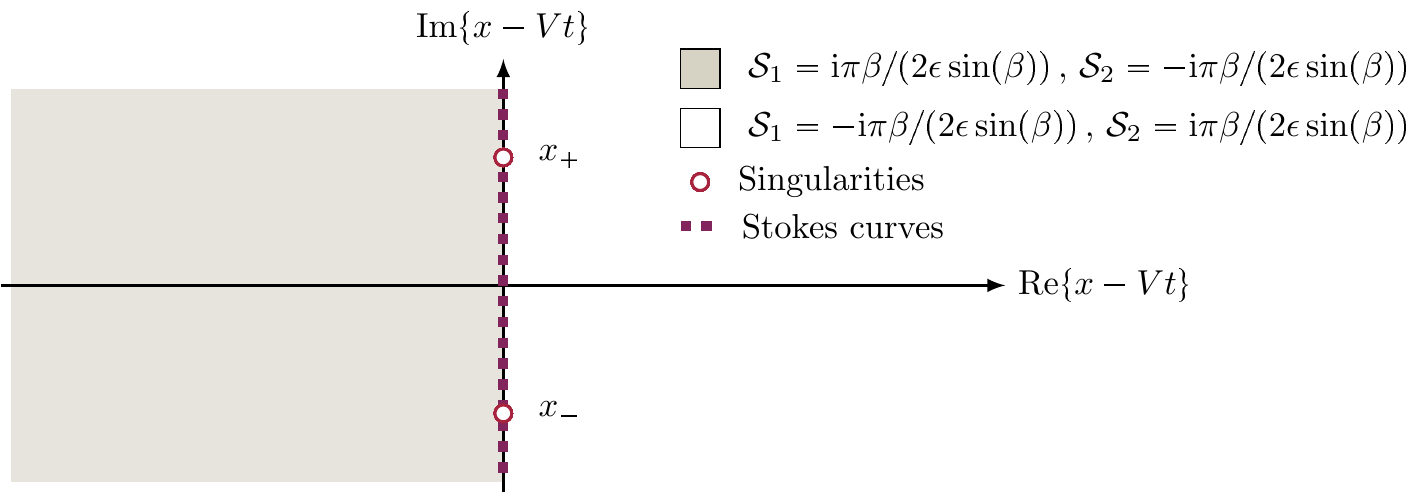}
	\caption{Stokes structure for the CDK2 equation that is associated with the leading-order focusing solitary wave \eqref{LESNLS}. {Both} Stokes curves follow the dashed line (so they overlap). They are generated by the singularities $x = V t \pm \i\pi/(2A)$ (which we show as red circles) of the leading-order solution. The Stokes multipliers $\mathcal{S}_1$ and $\mathcal{S}_2$ take different values on the two sides of the Stokes curves.
	We use shading to distinguish between the two regions.
	}
	\label{stokmult22}
\end{figure}

The singulants determine the Stokes structure of a solution. As we indicated in equation \eqref{STOKESCOND}, Stokes curves satisfy $\Im\{\chi\}=0$ and $\Re\{\chi\}>0$. Each singularity produces a Stokes curve along the line $\Re\{x\}=Vt$. {In Figure \ref{stokmult22}, we show a schematic illustration of the Stokes structure of {solutions of \eqref{TSIN} with leading-order behavior \eqref{LESNLS}}.}
 This figure also shows the Stokes multipliers (which we compute in Appendix \ref{bapkar2}), which take different values on the two sides of the Stokes curves.

\subsubsection{Prefactor Equation}\label{ch5pref}

Matching at the next-to-leading order (i.e., matching terms of size $\mathcal{O}(\psi_{j+1/2})$ as $j\rightarrow\infty$) in \eqref{LOTCD} gives the prefactor equation
\begin{equation}\label{cdpreffan}
	\frac{\i\Psi}{\sigma}\sum_{r=0}^j\frac{\left(\sigma \chi_t \right)^{2r+1}}{(2r+1)!} + \Psi_x\left(\left(1-4\lambda\right)\sum_{r=0}^j \frac{\chi_x^{2r+1}}{(2r+1)!} + 2\lambda \sum_{r=0}^j \frac{(2\chi_x)^{2r+1}}{(2r+1)!}\right) = 0 \, . 
\end{equation}
We again allow the upper bound of the sum to become infinite; this approximation introduces only an exponentially small error into the prefactor. We then obtain
\begin{equation}\label{cdpref2} 
	-\frac{\i\Psi}{\sigma}\sinh\left(\sigma V\chi_x\right) + \Psi_x\sinh\left(\chi_x\right)\left(1-4\lambda + 4\lambda\cosh\left(\chi_x\right)\right) = 0 \, . 
\end{equation}
For the first set of singulants in \eqref{2SINGS}, equation \eqref{cdpref2} implies that $\Psi = 0$. For the second set of singulants (which includes $\chi_1$ and $\chi_2$) in \eqref{2SINGS}, 
equation \eqref{cdpref2} simplifies to 
\begin{equation}\label{cdpreffa2} 
	\frac{\i\Psi}{\sigma}\sinh\left(\sigma V\chi_{x}\right) + {\Psi}_x\sinh\left(\chi_{x}\right) = 0 \, . 
\end{equation}
We solve equation \eqref{cdpreffa2} to obtain
\begin{equation}\label{CDPRE}
	\Psi(x,t) = f(t) \exp\left(-\frac{\i}{\sigma}\frac{\sinh(\sigma V\chi_{x})}{ \sinh(\chi_{x})}x\right) \,,	 
\end{equation}
where $f(t)$ is an arbitrary function that arises from integrating with respect to $x$. We denote the prefactor and arbitrary function that are associated with {$\chi_{\nu}$ for $\nu \in \{1,2\}$ by $\Psi_{\nu}$ and $f_{\nu}$, respectively}.

We determine $f(t)$ by matching the late-order terms with the leading-order solution near the singularities. (See Appendix \ref{CTPP} for the details of this analysis.) This yields the prefactor terms
\begin{equation} \label{cdinb}
	f_1(t) = \Lambda\mathrm{e}^{\i\left[\left(\frac{A^2-V^2}{2}\right)t +\left(V + \frac{\sin(\sigma V\beta({\lambda}))}{\sigma \sin(\beta({\lambda}))}\right)x_{+}\right]}\quad \text{and} \quad 	f_2(t) = \Lambda\mathrm{e}^{\i\left[\left(\frac{A^2-V^2}{2}\right)t +\left(V + \frac{\sin(\sigma V\beta({\lambda}))}{\sigma \sin(\beta({\lambda}))}\right)x_{-}\right]}\,, 
\end{equation}
where $\Lambda$ depends only on $\lambda$. We approximate the value of $\Lambda$ using a numerical computation. We show our approximation procedure in Appendix \ref{CTPP} for $\lambda = 1$. In this case, we obtain $\Lambda \approx 7.3$.

The strength of the singularity of the late-order terms \eqref{LOT} must be consistent with that of the leading-order behavior of the solitary-wave solution \eqref{LESNLS} as $x$ approaches the associated singularity. Recall that we are determining the late-order terms that arise from the singularities 
$x=x_+$ and $x=x_-$. 
The singularities of the leading-order solution \eqref{LESNLS} are poles of order $1$. For this behavior to be consistent with the late-order terms, we need $\gamma=1$. 
The complete form of the late-order terms is thus
\begin{align} \label{CDfullLOT}
	\psi_j &\sim \frac{\Lambda\exp\left(\i\left(\frac{(A^2-V^2)t}{2} +Vx_{+}-\frac{\sin(\sigma V\beta)}{\sigma \sin(\beta)}(x-x_{+})\right)\right) \Gamma(2j+1)}{(\i\beta(x-x_+))^{2j+1}}  \nonumber \\ & ~~~~~~~~~~~~~~~~~~~~ + \frac{\Lambda\exp\left(\i\left(\frac{(A^2-V^2)t}{2} +Vx_{-}-\frac{\sin(\sigma V\beta)}{\sigma \sin(\beta)}(x-x_{-})\right)\right) \Gamma(2j+1)}{(-\i\beta(x-x_-))^{2j+1}}      \quad \text{ as } \quad  j\rightarrow\infty\,. 
\end{align}


\subsection{Stokes Switching}

We now apply the matched asymptotic-expansion approach of \cite{olde1995stokes} to determine the form of the exponentially small oscillations. We give an outline of this method in Section \ref{eaback}, and we give the technical details in Appendix \ref{bapkar2}. From this analysis, we find that the asymptotic behavior of the exponentially small oscillations for $x \in \mathbb{R}$ in the limit $\epsilon\to 0$ are
\begin{equation}  \label{cdEXPTERSYM}
 	\psi_{\text{exp}}  \sim
 	\begin{dcases}
        \textcolor{white}{-}\frac{\i\pi\beta\Lambda}{\epsilon\sin(\beta)} \a^{-\frac{\beta\pi}{2A\epsilon}}\a^{\i\left(\left(A^2+V^2\right)\frac{t}{2}-\frac{\sin(\sigma V\beta)}{\sigma\sin(\beta)}(x-Vt)\right)}\mu(x,t)\,,& x < V t  \\
        -\frac{\i\pi\beta\Lambda}{\epsilon\sin(\beta)} \a^{-\frac{\beta\pi}{2A\epsilon}}\a^{\i\left(\left(A^2+V^2\right)\frac{t}{2}-\frac{\sin(\sigma V\beta)}{\sigma\sin(\beta)}(x-Vt)\right)}\mu(x,t)\,, & x > V t\,, \\
    \end{dcases}
\end{equation}
where
 \begin{equation}
 	\mu(x,t) = \sinh\left(\tfrac{\pi}{2A}\left(V+\tfrac{\sin(\sigma V\beta)}{\sigma\sin(\beta)}\right)\right)\cos\left(\tfrac{\beta}{\epsilon}(x-Vt)\right)  + \i\cosh\left(\tfrac{\pi}{2A}\left(V+\tfrac{\sin(\sigma V\beta)}{\sigma\sin(\beta)}\right)\right)\sin\left(\tfrac{\beta}{\epsilon}(x-Vt)\right)\,.
 \end{equation}
 The transition between the two asymptotic behaviors occurs in a narrow region of width $\mathcal{O}(\sqrt{\epsilon})$ around the Stokes curves.

\section{Families Of Discrete Central-Difference Karpman Equations} \label{ch4}

By comparing the results of Sections \ref{TKE} and \ref{ch3}, we see that the discretization changes both the form of the exponentially small oscillations in a solution and the critical value of $\lambda$ at which such oscillations appear. To understand such changes, we examine two families of {discrete} Karpman equations. 

We fix the order of the finite-difference approximation of \cjl{\eqref{CDNLSGHO}} 
and study the effect of discretization on solutions of a family of 
{discrete} higher-order Karpman equations. We obtain these {discrete} equations by scaling lattice {equations, and {we} then study the resulting advance--delay equations using exponential asymptotics.} This allows us to compare solutions of each {discrete} Karpman equation with those of the associated {continuous} higher-order Karpman equation \eqref{CDNLSGHO} (see Section \ref{OLOS}). We  then determine how the order of the equation affects the critical value of $\lambda$.

We then consider different discretizations of the continuous fourth-order Karpman equation \eqref{KARP} to determine whether or not increasing the order of the finite-difference discretization produces solution behavior that tends to the behavior that we observed for the continuous system in Section \ref{TKE}.


\subsection{{Discrete} Higher-Order Karpman Equations} \label{AFOHOD}

In Section \ref{OLOS}, we studied solutions of the continuous higher-order  Karpman equation \eqref{CDNLSGHO} and determined when they have nanopteron solutions. We now discretize these equations using a second-order central-difference discretization to study how the critical value of $\lambda$ changes as a result of discretization. We consider both odd and even values of $k$ in equation \eqref{CDNLSGHO}. One can calculate the finite-difference coefficients for odd values of $k$, but their general form is more complicated than the coefficients for even values of $k$. However, although those coefficients have a more complicated form, the analysis follows the same steps. Therefore, we present explicit calculations for even values of $k$ and present only the results of such calculations for odd values of $k$.

\begin{figure}[tb]
	\centering
	\includegraphics{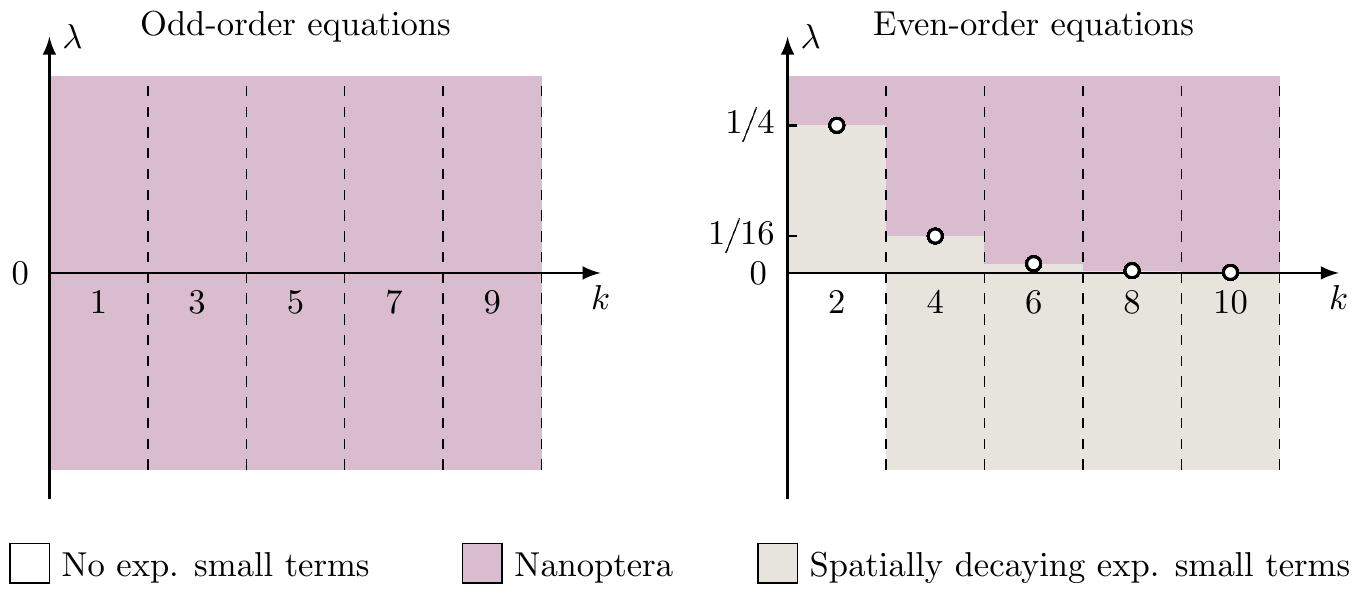}
	\caption{Bifurcation diagram of the traveling-wave solutions of the second-order central-difference {discrete} higher-order Karpman equations \eqref{GKTSIN} for different values of $k$ and $\lambda$, where the order of the equation is $k+2$. If $k$ is odd, nanoptera exist for all $\lambda$. If $k$ is even, nanoptera exist only when $\lambda > 1/2^k$; there is a bifurcation at $\lambda = 1/2^k$ for all even $k$. The circles
	mark the bifurcation points. 
	}
	\label{gswbehav2}
\end{figure}

\textcolor{black}{A lattice version of the continuous higher-order Karpman equation \eqref{CDNLSGHO} with even $k$ is 
\begin{align}\nonumber
\frac{\i}{2\tau}(w_{m,n+1} - w_{m,n-1}) &+ \frac{1}{2h^2}(w_{m+1,n} - 2w_{m,n} + w_{m-1,n}) \\
& ~~~~~~~~~~~~+ |w_{m,n}|^2w_{m,n}  + \frac{\i^k \lambda\epsilon^2}{2h^4}\sum_{l=0}^{k+2}  (-1)^l\binom{k+2}{l}w_{m+k/2+1-l,n}  = 0 \,.
\end{align}
We apply a spatial scaling of $x = hm$ and a temporal scaling of $t = \tau n$, and we then let ${w}_{m,n} = \psi(x,t)$ to obtain {an advance--delay equation. We then set $h = \epsilon$ and $\tau = \sigma \epsilon$ to obtain}}
\begin{align} \label{GKTSIN}
	\i &\frac{\psi(x,t+\sigma\epsilon)-\psi(x,t-\sigma\epsilon)}{2\sigma\epsilon} + \frac{1}{2}\frac{\psi(x+\epsilon,t)-2\psi(x,t)+\psi(x-\epsilon,t)}{\epsilon^2}  \nonumber \\ & ~~~~~~~~~~~~~~~~~~~~~~~~~~~~~ - \frac{\i^{k}\lambda}{2\epsilon^{2}}\sum_{l=0}^{k+2} (-1)^l\binom{k+2}{l}\psi(x + [k/2+1-l]\epsilon,t) + |\psi(x,t)|^2\psi(x,t) = 0\,.	
\end{align}
\textcolor{black}{One can obtain \eqref{GKTSIN} directly by applying a second-order central difference discretization to the  continuous higher-order  Karpman equation \eqref{CDNLSGHO}}. 
Applying a Taylor-series approximation about $\epsilon=0$ yields the infinite-order partial differential equation
\begin{align} \label{GKCDNLS3}
	&\i \sum_{r=0}^{\infty} \frac{(\sigma \epsilon)^{2r}}{(2r+1)!}\frac{\partial^{2r+1}}{\partial t^{2r+1}}\psi(x,t) + \sum_{r=0}^{\infty} \frac{\epsilon^{2r}}{(2r+2)!} 
\frac{\partial^{2r+2}}{\partial x^{2r+2}}\psi(x,t) + |\psi(x,t)|^2\psi(x,t) \nonumber \\& ~~~~~~~~~~~~~~~~~~~~~~~~~~~~~~~~~~~~ - \i^{k}\lambda\sum\limits_{l=0}^{k+2} (-1)^l\binom{k+2}{l} \sum_{r=0}^{\infty} \frac{(k/2+1-l)^{2r+2}\epsilon^{2r}}{(2r+2)!} 
\frac{\partial^{2r+2}}{\partial x^{2r+2}}\psi(x,t)  = 0\,. 
\end{align}
We follow the same method as in Section \ref{cdfase} and expand $\psi$ as an asymptotic power series in $\epsilon^{2}$. The leading-order solution $\psi_0$ satisfies the cubic NLS equation \eqref{GNLS}, and we again take $\psi_0$ to be the focusing solitary-wave solution \eqref{LESNLS}. 
Matching terms of size $\mathcal{O}(\epsilon^{2j})$ as $\epsilon \rightarrow 0$ in \eqref{GKCDNLS3} produces a recurrence relation for the subsequent terms in the asymptotic series. We determine the late-order terms as in Section \ref{cdLOT} using the late-order ansatz \eqref{LOT} to obtain the singulant equation
\begin{equation}\label{gksing}
\cosh(\chi_x)-1 + \i^{k+1}\lambda\sum\limits_{l=0}^{k/2} (-1)^l \binom{k+2}{l}(\cosh([k/2+1-l]\chi_x)-1) = 0 \, . 
\end{equation}
Solving \eqref{gksing} yields $k+2$ values of $\chi_x$.

Nanopteron solutions of {\eqref{GKTSIN}} 
exist if {and only if} there are {purely} imaginary values of $\chi_x$. 
There is a critical value $\lambda_c$ of $\lambda$ that determines whether or not such values exist. By solving \eqref{gksing}, we find that nanoptera occur for $\lambda > \lambda_c$ and that the critical value is 
\begin{equation} \label{gksingcond}
	\lambda_c = \frac{2}{\i^{k+1}\sum\limits_{l=0}^{k/2} (-1)^l \binom{k+2}{l}(\cosh(\pm\i\pi[k/2+1-l])-1)} = \frac{1}{2^k}\,. 
\end{equation}

When $\lambda>1/2^k$, there are imaginary {values of $\chi_x$ that satisfy equation \eqref{gksing}}; the associated traveling-wave solutions of \eqref{GKTSIN} are nanoptera.  {For $k\geq4$, the nanopteron solutions of \eqref{GKTSIN} have radiatively decaying exponential contributions in addition to the non-decaying oscillations.} When $ \lambda < 1/2^k$, all exponentially small contributions of \eqref{GKTSIN} are radiatively decaying; one thus obtains exponentially localized waves. We summarize these results for even-order equations in Figure \ref{gswbehav2}.

When $k$ is odd, the expressions for the central-difference coefficients are more complicated than in \eqref{GKTSIN}. By a direct calculation, one can show that at least one of the $k+2$ values of $\chi_x$ is always imaginary. Therefore, second-order central-difference {discrete} higher-order Karpman equations of odd order have nanopteron solutions for all $\lambda$. We summarize these results for odd-order equations in Figure \ref{gswbehav2}.

By comparing the bifurcation values for continuous \eqref{CDNLSGHO} and {discrete} \eqref{GKTSIN} higher-order Karpman equations, we see that critical value of $\lambda$ that permits nanopteron solutions is different for the continuous and {discrete} equations if the order of the equation is even. In the continuous equations, nanoptera solutions exist when $\lambda > 0$ (see Figure \ref{gswbehav}). However, for the {discrete} equations, nanoptera exist only when $\lambda > 1/2^k$ (see Figure \ref{gswbehav2}). If $k$ is odd, the continuous Karpman equation does not have such a bifurcation point. This is also true for our second-order central-difference {discrete} higher-order Karpman equation.


\subsection{Higher-Order Finite-Difference Discretization of the {Continuous} Fourth-Order Karpman Equation} \label{HOCDKE}

The exponentially small oscillations \eqref{EXPTERSYMJ} in nanopteron solutions of the continuous fourth-order Karpman equation \eqref{KARP} are different from the oscillations \eqref{cdEXPTERSYM} in solutions of the second-order central-difference fourth-order Karpman equation \eqref{TSIN}, even for solutions in which the leading-order solitary-wave is given by the same expression (e.g., the focusing solitary wave \eqref{LESNLS}). Motivated by this observation, we study the effect of discretization on the behavior of the oscillatory tails in nanopteron solutions of \eqref{KARP} as we increase the order of the discretization.

We gave the form of the exponentially small contributions of the nanoptera in \eqref{earemint}. The singulant, prefactor, and Stokes multiplier determine the asymptotic behavior of the oscillatory tails in these nanoptera. Therefore, the effect of discretization on a nanopteron depends on how the singulants, prefactors, and Stokes multipliers change when we discretize a system.

\begin{table}[tb]
    \centering
    \footnotesize
    \begin{tabular}{|c|c|c|c|c|c|c|c|c|c|c|}
    \cline{3-11}
    \multicolumn{2}{c}{} & \multicolumn{9}{|c|}{Stencil Term ($r$)} \\ \hline
        Derivative ($p$) & Accuracy ($2q$) & $-4$ & $-3$ & $-2$ & $-1$ & $0$ & $1$ & $2$ & $3$ & $4$ \\ \hline
         & $2$ & & & & $-1/2$ &  & $1/2$ & & & \\ \cline{2-11}
         $1$ & $4$ & & & $1/12$ & $-2/3$ &  & $2/3$ & $-1/12$ & &   \\ \cline{2-11}
         & $6$ & & $-1/60$ & $3/20$ & $-3/4$ &  & $3/4$ & $-3/20$ & $1/60$ & \\
         \hline
         & $2$ & & & & $1$ & $-2$ & $1$ & & & \\ \cline{2-11}
         $2$ & $4$ & & & $-1/12$ & $4/3$ & $-5/2$ & $4/3$ & $-1/12$ & &   \\ \cline{2-11}
         & $6$ & & $1/90$ & $-3/20$ & $3/2$ & $-49/18$ & $3/2$ & $-3/20$ & $1/90$ & \\
         \hline
         & $2$ & & & $1$ & $-4$ & $6$ & $-4$ & $1$ & & \\ \cline{2-11}
         $4$ & $4$ & & $-1/6$ & $2$ & $-13/2$ & $28/3$ & $-13/2$ & $2$ & $-1/6$ &  \\ \cline{2-11}
         & $6$ & $7/240$ & $-2/5$ & $169/60$ & $-122/15$ & $91/8$ & $-122/15$ & $169/60$ & $-2/5$ & $7/240$ \\\hline
    \end{tabular}
    \caption{Finite-difference coefficents $C_{p,2q,r}$ of the $r^{\mathrm{th}}$ stencil term in the central-difference approximation of a $p^{\mathrm{th}}$ derivative with an accuracy order of $2q$.
    }
    \label{fdtab}
\end{table}

Consider an advance--delay equation {that we obtain} from \eqref{KARP} by applying a finite-difference discretization with {the} spatial step size $\epsilon$ {and the temporal step size $\sigma \epsilon$}. 
We approximate the derivative terms using central-difference approximations with an accuracy order of $2q$, where $q$ is a positive integer.
 We obtain the central-difference approximation using a ``stencil'', which gives an approximation of the derivative terms. The stencil for finding partial derivatives of a function {$g$} with respect to $x$ is
\begin{equation}
	\pdiff{^p {g}}{x^p} \approx \frac{1}{\epsilon^p} \sum_{r=-q}^{q} C_{p,2q,r} {g}(x+ r \epsilon)\,,
\end{equation}

The ``CDKA'' version of the fourth-order Karpman equation \eqref{KARP} is
\begin{align} \label{ckda}
		\frac{\i}{\sigma\epsilon}\sum\limits_{r=-q}^{q} C_{1,2q,r}\psi(x,t+r\sigma\epsilon) & + \frac{1}{2\epsilon^2}\sum\limits_{r=-q}^{q} C_{2,2q,r}\psi(x+r\epsilon,t)\nonumber\\ & + \frac{\lambda}{2\epsilon^2}\sum\limits_{r=-(q+1)}^{q+1} C_{4,2,r}\psi(x+r\epsilon,t) + \left|\psi(x,t)\right|^2\psi(x,t) = 0\,.
\end{align}

We expand \eqref{ckda} in the limit $\epsilon \rightarrow 0$ to obtain
\begin{align} \label{fineq}
		 2\i\sum\limits_{r=1}^{q} C_{1,2q,r}\sum\limits_{k=0}^\infty \frac{r^{2k+1}(\sigma\epsilon)^{2k}}{(2k+1)!} \frac{\partial^{2k+1}}{\partial t^{2k+1}} \psi &+ \sum\limits_{r=1}^{q} C_{2,2q,r}\sum\limits_{k=0}^\infty \frac{r^{2k+2}\epsilon^{2k}}{(2k+2)!} \frac{\partial^{2k+2}}{\partial x^{2k+2}} \psi \nonumber\\  & + \lambda \sum\limits_{r=1}^{q+1} C_{4,2q,r}\sum\limits_{k=0}^\infty \frac{r^{2k+2}\epsilon^{2k}}{(2k+2)!} \frac{\partial^{2k+2}}{\partial x^{2k+2}} \psi +\left|\psi\right|^2\psi = 0\,. 
\end{align}
Following the method in Section \ref{cdfase}, we expand $\psi$ as an asymptotic power series in $\epsilon^{2}$. The leading-order solution $\psi_0$ satisfies the cubic NLS equation \eqref{GNLS}, and we again take $\psi_0$ to be the focusing solitary-wave solution \eqref{LESNLS}. Matching terms of size $\mathcal{O}(\epsilon^{2j})$ as $\epsilon\to 0$ in \eqref{fineq} yields a recurrence relation for the subsequent terms in the asymptotic series. 
As in Section \ref{cdLOT}, we determine the late-order terms using the ansatz \eqref{LOT}. 


\subsubsection{Singulants}

The singulant equation is
\begin{equation}\label{gsing}
	\sum\limits_{r=1}^{q} C_{2,2q,r}(\cosh(r\chi_x)-1) + \lambda \sum\limits_{r=1}^{q+1} C_{4,2q,r}(\cosh(r\chi_x)-1) = 0 \, .
\end{equation}
It is satisfied by imaginary values of $\chi_x$ when
\begin{equation} \label{cdgksingcond}
	\lambda \geq \lambda_c = \frac{\sum\limits_{r=1}^{q} C_{2,2q,r}(1-(-1)^r)}{\sum\limits_{r=1}^{q+1} C_{4,2q,r}((-1)^r-1)} \,. 
\end{equation}
Therefore, when $\lambda>\lambda_c$, traveling-wave solutions of \eqref{ckda} are nanoptera. When $\lambda < \lambda_c$, there are no imaginary values of $\chi_x$, so all exponentially small contributions to traveling-wave solutions of \eqref{ckda} decay radiatively; {the traveling-wave} solutions are exponentially localized in space. 

\begin{figure}[tb]
\centering
\includegraphics{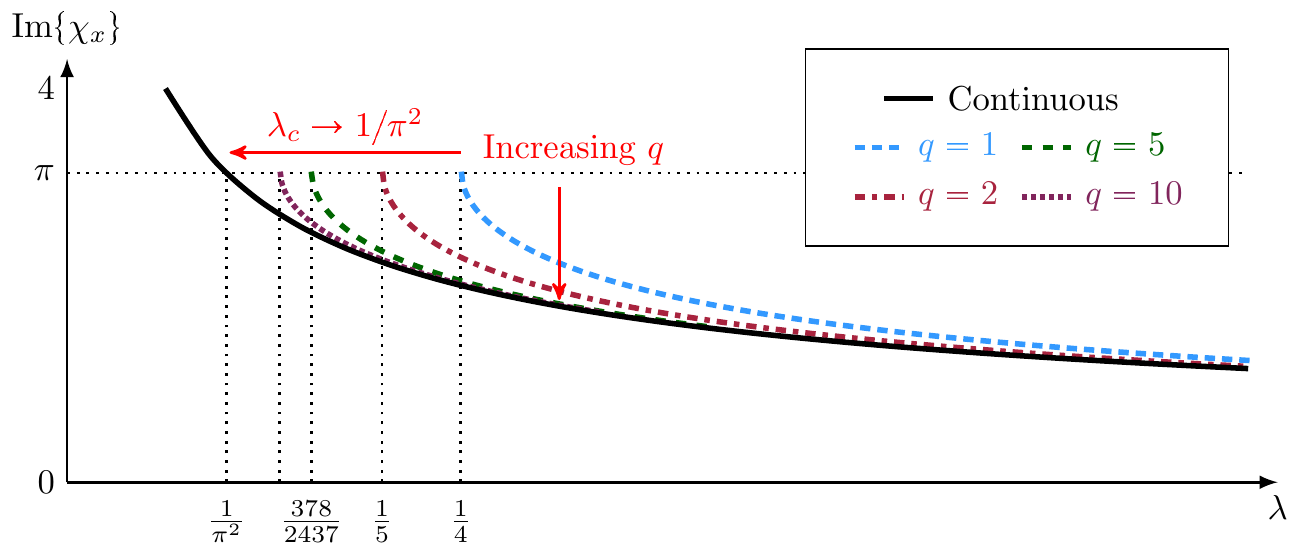}
\caption{A comparison of the singulants for several central-difference discretizations of the {continuous} fourth-order Karpman equation \eqref{KARP} with central-difference discretizations with accuracy orders of $2q$ for different choices of $q$. {We include the singulant of the continuous fourth-order Karpman equation in the figure for reference.}. 
As the accuracy order $2q$ becomes large, the critical value $\lambda_c$ for the existence of nanoptera approaches $1/\pi^2$. When $\lambda > \lambda_c$, the behavior of the discrete singulant approaches that of the singulant of the continuous equation as one increases $q$.	
	}
	\label{singcomp}
\end{figure}

When $\lambda > \lambda_c$, there are two imaginary values of $\chi_x$; they differ only in their sign. 
In Figure \ref{singcomp}, we show the value of $\chi_x$ with a positive imaginary part as a function of $\lambda$ for the continuous and {discrete} fourth-order Karpman equations. From this figure, we make two important observations. First, for singulants of the discrete equation, \change{we observe that the critical value $\lambda_c$ approaches $1/\pi^2$ as we increase the order of the discretization. We further support this observation by directly 
calculating $\chi_x$ using \eqref{gsing} for larger values of $q$. (For readability, we omit these values from Figure \ref{singcomp}.)} This differs from the continuous case, for which $\lambda_c = 0$. Second, when $\lambda > \lambda_c$, the singulant behavior of the {discrete} fourth-order Karpman {equation} \eqref{ckda} tends to the singulant behavior {of the continuous fourth-order Karpman equation \eqref{KARP}} as we increase the order of the discretization. 

These observations demonstrate that the change in the critical value of $\lambda$ is an unavoidable result of discretization that persists even when one increases the accuracy of a central-difference approximation. However, when $\lambda > \lambda_c$, one recovers the singulant behavior of the continuous fourth-order Karpman equation \eqref{KARP} in the limit of large finite-difference accuracy order.


\subsubsection{{Prefactors}}

\begin{figure}[tb]
\centering
\includegraphics{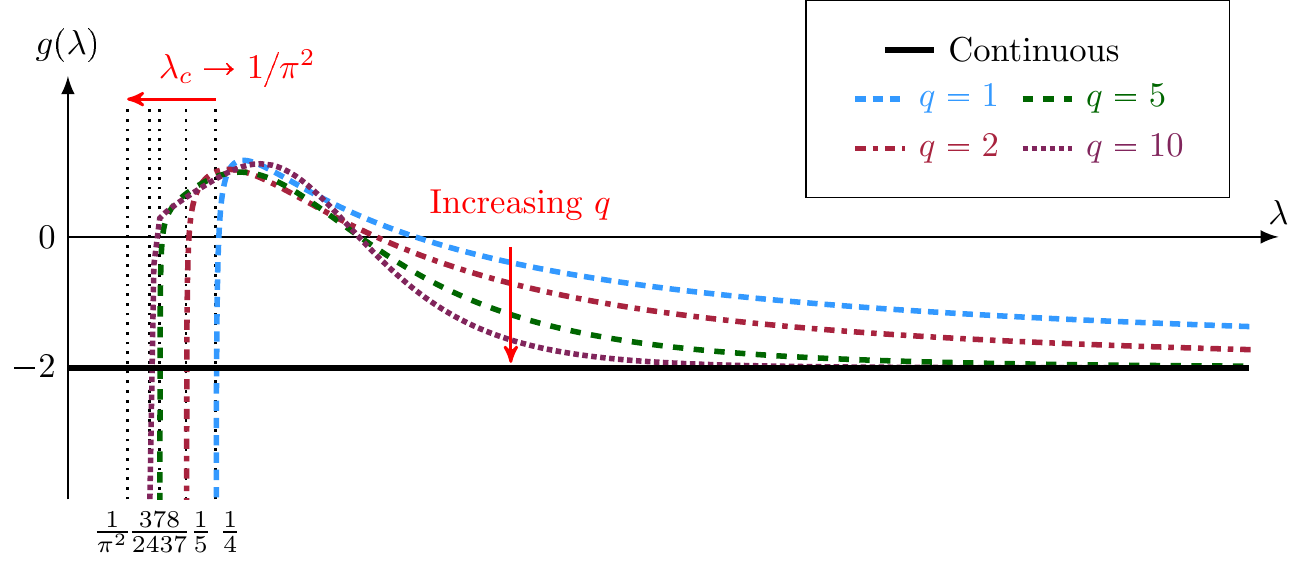}
	\caption{A comparison of the prefactor term ${g(\lambda)}$ for several central-difference discretizations of the continuous fourth-order Karpman equation \eqref{KARP} with different discretization accuracy orders $2q$, a speed of $V = 2$, and $\sigma=1.1$, where $\sigma$ is the ratio of the spatial discretization step to the temporal discretization step. We also show the prefactor term for the continuous fourth-order Karpman equation. As we increase the order of the central-difference approximation, when $\lambda>\lambda_c$, {the functional form of the prefactor for our {discrete} fourth-order Karpman equation approaches the functional form of the prefactor (see the solid black line) for the continuous fourth-order Karpman equation. As $\lambda$ approaches the critical value $\lambda_c$ from above, ${g(\lambda)}$ first increases and then decreases, with ${g(\lambda)} \to -\infty$ as $\lambda\to\lambda_c$.}}
	\label{prefcomp}
\end{figure}

The prefactor equation is
\begin{equation}\label{6.2 PF}
	\frac{2\i\Psi}{\sigma}\sum\limits_{r=1}^{q}C_{1,2q,r}\sinh(r\sigma\chi_t) + \Psi_x\left(\sum\limits_{r=1}^{q}rC_{2,2q,r}\sinh(r\chi_x)+\lambda\sum\limits_{r=1}^{q+1}rC_{4,2q,r}\sinh(r\chi_x)\right) = 0 \, .
\end{equation}
When $\lambda>\lambda_c$, the prefactor that is associated with the singularities 
$x_{\pm}$ has the form 
\begin{equation} \label{eq:FDpref}
    \Psi(x,t) = \Lambda(\lambda)\a^{\i((A^2-V^2)t/2 + {g(\lambda)}(x - x_{\pm}) + V x_{\pm})}\,,
\end{equation}
where ${g(\lambda)}$ and $\Lambda(\lambda)$ are the only parts of the expression in \eqref{eq:FDpref} that depend on the discretization. The prefactor of the continuous fourth-order Karpman equation \eqref{KARP} also has the form in \eqref{eq:FDpref}. By comparing the expression in \eqref{eq:FDpref} with \eqref{fullLOT}, we see that ${g(\lambda)} = -V$ in the prefactors of the series terms {for}
the continuous fourth-order Karpman equation. 
The function ${g(\lambda)}$ in the prefactors of the late-order terms for the {discrete} Karpman equation \eqref{ckda} has the form
\begin{equation}
    {g(\lambda)} = -\frac{2\sum\limits_{r=1}^{q}C_{1,2q,r}\sinh(r\sigma V\chi_x)}{\sigma\left(\sum\limits_{r=1}^{q}rC_{2,2q,r}\sinh(r\chi_x)+\lambda\sum\limits_{r=1}^{q+1}rC_{4,2q,r}\sinh(r\chi_x)\right)} \,.
\end{equation}
In Figure \ref{prefcomp}, we compare ${g(\lambda)}$ for the continuous and lattice Karpman equations with $V = 2$.

When $\lambda > \lambda_c$, we see that ${g(\lambda)} \to -V$ as $q$ increases; this is the value of ${g(\lambda)}$ for the continuous fourth-order Karpman equation. As $\lambda$ approaches $\lambda_c$ from above, ${g(\lambda)}$ moves away from $-V$ before decreasing, with ${g(\lambda)} \to \pm\infty$ as $\lambda \to \lambda_c$, where the sign depends on {the value of} $\sigma$ and the sign of $V$. We show an example of this behavior in Figure \ref{prefcomp}. As $q\to\infty$, this limiting behavior occurs in a progressively narrower region. Therefore, when $\lambda > \lambda_c$, the behavior of the prefactor for our {discrete} fourth-order Karpman {equation \eqref{fineq}} tends to the behavior of the prefactor for the continuous fourth-order Karpman equation as we increase the order of the finite difference.

One can show that $\Lambda(\lambda)$ in \eqref{eq:FDpref} tends to the corresponding value for the continuous equation \eqref{KARP} by following the same steps as in the inner analysis in Appendix \ref{CTPP}. One can also show that increasing the order $2q$ of the approximation removes terms {that appear in the inner expansion 
{that we use for the analysis in} Appendix \ref{CTPP}}.
 In the limit $q \rightarrow \infty$, the remaining terms {in the inner expansion} are exactly the terms from the inner analysis of the continuous fourth-order Karpman equation (see Appendix \ref{CTP}). Therefore, when $\lambda > \lambda_c$, the prefactor of the late-order terms in the {discrete} Karpman equation \eqref{ckda} tend to the prefactor of the late-order terms in the continuous fourth-order Karpman equation \eqref{KARP} as we increase the order of the discretization.


\subsubsection{Exponentially small terms}

Using a similar exponential asymptotic analysis {as} the one in Appendix \ref{bapkar2}, one can show that if $\lambda > \lambda_c$, then {the Stokes multipliers} $\mathcal{S}$ for our {discrete} Karpman equation {approach} the analogous {expressions} for the continuous fourth-order Karpman equation as $q\to\infty$. 
Therefore, when $\lambda > \lambda_c$, the late-order terms and the exponentially small oscillations tend to the continuous results in \eqref{fullLOT} and \eqref{CONS}{, respectively, as $q\to\infty$.}

The results of {the analysis of the discrete {Karpman} equation \eqref{ckda}} allow us to draw the following conclusions:
\begin{itemize}
    \item{{In} our {discrete} Karpman equations, there is a nonzero critical value $\lambda_c$ (which we showed in \eqref{cdgksingcond}) of $\lambda$ that approaches
    $1/\pi^2$ as we increase the order of the discretization. It is impossible to obtain nanoptera when $\lambda < 1/\pi^2$ even for large values of $q$.} 
    \item{As we showed in Figures \ref{singcomp} and \ref{prefcomp}, if $\lambda > \lambda_c$, then the late-order terms and the associated exponentially small oscillations tend to the corresponding continuous behavior as $q \to \infty$.}  
\end{itemize}


\section{Conclusions and Discussion} \label{sec7}

We examined nanopteron solutions of continuous and {discrete} Karpman equations, and we investigated the effects of discretization on such solutions. We studied the asymptotic behavior of the exponentially small oscillations in both situations, and we determined that discretization alters both the oscillatory behavior and the critical parameter values at which oscillations appear.

Our analysis of Karpman equations used exponential asymptotic methods that were developed in \cite{chapman1998exponential,olde1995stokes,joshi2015stokes,joshi2017stokes,joshi2019generalized,king2001asymptotics}. We found that the solutions of these equations have Stokes curves that produce exponentially small contributions in the solutions and thus result in GSWs. If $\lambda$ is at least
certain critical values, these GSWs are nanoptera because the amplitude of the oscillatory tails do not decay in space.

We first studied nanopteron solutions of the continuous fourth-order Karpman equation \eqref{KARP}. (Stationary nanopteron solutions of this equation were studied previously in \cite{karpman1994solitons}.) When $\lambda > 0$, we found that the solution has two Stokes curves, which produce non-decaying oscillations; therefore, the traveling-wave solution {is a nanopteron}. 
We analyzed the asymptotic behavior of these oscillatory tails \eqref{EXPTERSYMJ}, and we verified our asymptotic results using numerical computations. We found that the stationary nanoptera agree with the results in \cite{karpman1994solitons} and that traveling nanoptera have a slowly varying {periodic amplitude}, which is consistent with the numerical computations in \cite{karpman1999evolution}. 

We then considered {solutions {of} the} continuous higher-order Karpman equation \eqref{CDNLSGHO}. We found that if the highest-order derivative is odd, then traveling-wave solutions are always nanoptera. By contrast, if the highest-order derivative is even, we found that traveling-wave solutions are nanoptera only when $\lambda > 0$. Our results for the third-order derivative are consistent with previous studies of the continuous third-order Karpman equation \cite{karpman1993stationary,karpman1993radiation,wai1990radiations,wai1986nonlinear}. We are not aware of studies of nanoptera in continuous Karpman equations of odd order of more than three.

We then studied a second-order central-difference approximation \eqref{TSIN} of the {continuous} fourth-order Karpman equation \eqref{KARP}. {We obtained this equation by scaling a lattice version of \eqref{KARP}}. When $\lambda < 0$, we found that solutions of \eqref{TSIN} have no exponentially small oscillations. When $\lambda \in (0,1/4)$, we found that solutions of \eqref{TSIN} are GSWs that decay radiatively in space, so they are exponentially localized. When $\lambda > 1/4$, we found that solutions of \eqref{TSIN} are nanoptera. We then derived an asymptotic description \eqref{cdEXPTERSYM} of these nanoptera and found that the oscillations have different amplitudes and periodicity than the nanopteron solutions of \eqref{KARP}. These results demonstrate that discretization alters both the critical value of the bifurcation parameter and the behavior of the exponentially small oscillations. 

We found similar results for second-order central-difference discretizations of higher-order Karpman equations. {In particular, discretizing these systems also alters the critical value of the bifurcation parameter}. If the highest-order derivative is even, the critical value of $\lambda$ has a nonzero positive value {in $(0,1/4)$}. This differs from the situation for continuous {higher-order  Karpman equations whose highest-order derivative is even}{; these equations} have a bifurcation at $\lambda = 0$. We summarize these results in Table \ref{restab1}.

\begin{table}[tb]
    \centering
    \footnotesize
    \begin{tabular}{|c?c|c|}
   \hline
        Order of Highest Derivative & Continuous Equation & {Discrete} Equation \\ \thickhline
        Odd & Always nanoptera & Always nanoptera  \\ \hline
        Even (order = $k+2$) & Nanoptera for $\lambda > 0$ & Nanoptera for $\lambda \geq 1/2^k$  \\ \hline
    \end{tabular}
    \caption{Comparison of the bifurcation values for nanoptera in continuous and {discrete} higher-order  Karpman equations. Discretization changes the critical value of $\lambda$ for Karpman equations if the highest derivative is even, but it does not change the critical value if the highest derivative is odd.}
    \label{restab1}
\end{table}

We studied the effects of discretization on nanoptera using arbitrary-order central-difference discretizations of the continuous fourth-order Karpman equation \eqref{KARP}. We summarize these effects in Table \ref{restab2}. Notably, we found that nanoptera are not possible in {our discrete} Karpman equations when $\lambda < 1/\pi^2$ for any order of the central-difference discretization that we used to obtain the {discretized} {equations}. Consequently, all traveling waves in {our discrete} Karpman equations with $\lambda < 1/\pi^2$ must be exponentially localized in space. 

\textcolor{black}{As we noted in Section \ref{GSCH}, the existence of critical values of $\lambda$ {that separate nanopteron and solitary-wave solutions} is analogous to the Peierls--Nabarro energy barrier, which limits the existence of traveling kink solutions in certain lattice systems \cite{aigner2003new,currie1980statistical,peyrard1984kink}. 
{For example, certain lattice systems only support kinks whose speed exceeds some critical value.} 
Kinks that travel slower than this critical speed generate radiation, and they thus eventually slow down and stop. Our {traveling-wave solutions of \eqref{fineq}} for $\lambda \in (0, \lambda_c]$ are traveling solitary waves; they correspond to traveling kink solutions. When $\lambda > \lambda_c$, {traveling-wave solutions of \eqref{fineq} are one-sided nanoptera, which radiate energy as exponentially small oscillations}. 
Two-sided nanopteron solutions {of \eqref{fineq}} are possible;
they correspond to stable solutions on an {oscillatory} background. It is known that such solutions also exist in lattices with a Peierls--Nabarro barrier \cite{duncan1993solitons}. These similarities between {traveling-wave behavior in systems with a Peierls--Nabarro energy barrier and traveling-wave behavior in discrete Karpman equations} indicate that one can use exponential asymptotics to give insights into the Peierls--Nabarro barrier. 
The fact that $\lambda_c$ depends on the choice of discretization supports the observation in \cite{kevrekidis2002continuum} that the Peierls--Nabarro barrier is {not the same in
continuous wave equations and their discretizations.}
}

\begin{table}[tb]
    \centering
    \footnotesize
    \begin{tabular}{|c||c|c|c|c|c|}
    \cline{1-6}
    \multicolumn{1}{|c}{Type of Equation} & \multicolumn{5}{?c|}{Type of Traveling-Wave Solutions} \\ \thickhline
    \multicolumn{1}{|c}{} & \multicolumn{1}{?p{0.1\textwidth}}{\centering $\lambda < 0$} & \multicolumn{4}{|c|}{$\lambda > 0$} \\ \cline{2-6}
    \multicolumn{1}{|c}{Continuous Equation} & \multicolumn{1}{?c}{\centering No exp. small} & \multicolumn{4}{|c|}{\centering Nanoptera:} \\ 
    \multicolumn{1}{|c}{} & \multicolumn{1}{?c}{oscillations} & \multicolumn{4}{|c|}{$\psi_{\mathrm{exp,\,cont.}} \sim \mathcal{S} \Psi \mathrm{e}^{-\chi/\epsilon}$}  \\ \thickhline
     \multicolumn{1}{|c}{{Discrete} Equation} & \multicolumn{2}{?p{0.2\textwidth}}{\centering $\lambda \leq 1/\pi^2$} & \multicolumn{1}{|p{0.13\textwidth}}{\centering $1/\pi^2 < \lambda < \lambda_c$} & \multicolumn{1}{|p{0.13\textwidth}}{\centering $\lambda_c \leq \lambda \leq 1/4$} & \multicolumn{1}{|p{0.13\textwidth}|}{\centering  $1/4 < \lambda$} \\ \cline{2-6}
    \multicolumn{1}{|c}{{Central}-Difference Order: $2q$} & \multicolumn{3}{?c}{No exponentially small} &  \multicolumn{2}{|c|}{Nanoptera:} \\ 
    \multicolumn{1}{|c}{$\lambda_c \to 1/\pi^2$ as $q \to \infty$} & \multicolumn{3}{?c}{oscillations} &  \multicolumn{2}{|c|}{$\psi_{\mathrm{exp,\,disc.}}\to \psi_{\mathrm{exp,\,cont.}}$ as $q \to\infty$}  \\ \hline
    \end{tabular}
    \caption{Comparison of traveling-wave solutions of the {continuous fourth-order } Karpman equation \eqref{KARP} and the {discrete fourth-order } Karpman equation \eqref{ckda} with a central-difference discretization order of $2q$. The continuous Karpman equation has nanoptera when $\lambda > 0$, and the behavior of exponentially small oscillations of its solutions is given by terms of the form $\psi_{\mathrm{exp,\,cont.}}$. The {discrete Karpman equation} has nanoptera when $\lambda > \lambda_c$, where $1/\pi^2 < \lambda_c \leq 1/4$ and $\lambda_c \to 1/\pi^2$ as $q \to \infty$. When $\lambda > \lambda_c$, the behavior of its solution's exponentially small oscillations, which is given by $\psi_{\mathrm{exp,\,disc.}}$, tends to the continuous behavior $\psi_{\mathrm{exp,\,cont.}}$ as $q\to\infty$.      However, when $\lambda < 1/\pi^2$, the {discrete} Karpman equation cannot yield nanopteron solutions for any value of $q$. This is an unavoidable difference between traveling waves in the continuous and {discretized} Karpman equations.}

    \label{restab2}
\end{table}

\appendix

\section{Detailed Calculations for the Continuous Fourth-Order Karpman Equation}

We now give detailed calculations for obtaining the late-order terms and exponentially small contributions of the continuous fourth-order Karpman equation \eqref{KARP}.


\subsection{Local Analysis Near The Singularities} \label{CTP}

We match the late-order terms \eqref{LOT} with the inner behavior near the singularities $x=x_+$ and $x=x_-$ to determine the time-dependent prefactor term $f(t)$ in \eqref{PREFF}. The asymptotic expansion \eqref{asyexp} fails to be asymptotic if $\epsilon^{2j}\psi_j \sim \epsilon^{2j+2}\psi_{j+1}$ as $\epsilon \to 0$; this occurs when $x-x_{\pm} = \mathcal{O}(\epsilon)$. We rescale the variables in \eqref{KARP} by setting $\epsilon\eta=x-x_{\pm}$ and $\phi(\eta,t)=\epsilon\psi(x,t)$. We retain the leading-order terms as $\epsilon\rightarrow0$; these terms have size $\mathcal{O}\left(\epsilon^{-3}\right)$. Near the singularities, the rescaled version of equation \eqref{KARP} becomes
  \begin{equation} \label{INNE}
	\frac{1}{2}\pdiff{^2\phi}{\eta^2}+\frac{\lambda}{2}\pdiff{^4\phi}{\eta^4} + \phi\overline{\phi}\phi = 0\,, 
\end{equation}
where the neglected terms are at most $\mathcal{O}\left(\epsilon^{-2}\right)$ as $\epsilon \rightarrow 0$. 

We cannot directly apply the inner matching from \cite{chapman1998exponential} to the inner expression \eqref{INNE} because the conjugation operation in \eqref{INNE} is not analytic. Instead, we define two new analytic functions, $U$ and $V$, with $U = \phi(x,t)$ and $V = \overline{\phi}(x,t)$ on $x \in \mathbb{R}$. This allows us to obtain a coupled system, whose solution we analytically continue away from the real axis.  

We conjugate \eqref{INNE} to obtain a second equation, and we then study the coupled system 
  \begin{equation} \label{INNEu}
	\frac{1}{2}\pdiff{^2U}{\eta^2}+\frac{\lambda}{2}\pdiff{^4 U}{\eta^4} + U^2V = 0\,, 
\end{equation}
  \begin{equation} \label{INNEv}
	\frac{1}{2}\pdiff{^2V}{\eta^2}+\frac{\lambda}{2}\pdiff{^4 V}{\eta^4} + UV^2 = 0\,. 
\end{equation}

We use the asymptotic expansions 
\begin{equation} \label{INEX}
	U(\eta,t) \sim \sum_{j=0}^\infty\frac{u_j(t)}{\eta^{2j+1}} \quad \text{as} \quad |\eta|\rightarrow\infty \quad\quad \text{and} \quad\quad V(\eta,t) = \sum_{j=0}^\infty\frac{v_j(t)}{\eta^{2j+1}} \quad \text{as} \quad |\eta|\rightarrow\infty\,. 
\end{equation}
To obtain $u_0$ and $v_0$, we need to determine the leading-order behavior (from \eqref{ILOAV}) of the outer solutions $\psi_0$ and $\overline{\psi}_0$ near the singularities.
The leading-order behavior of $\overline{\psi}_0$ near the singularities is 
\begin{equation} \label{ILOAVC} 
	\overline\psi_0(x,t) \sim \mp \frac{\i}{x-x_{\pm}}\mathrm{e}^{-\i\left(\left(A^2-V^2\right)t/2 +Vx_{\pm}\right)} \quad  \text{  as  } \quad  x\rightarrow x_{\pm}\,,
\end{equation}
where the upper and lower sign choices correspond.

Using Van Dyke's matching principle \cite{van1964perturbation}, we match \eqref{ILOAV} and \eqref{ILOAVC} in the inner limits as $x\rightarrow x_+$ and $x\rightarrow x_-$ with the inner expansions \eqref{INEX} in the outer limit $|\eta| \rightarrow \infty$. This gives the first terms in the series that match the leading-order behaviors of the inner and outer regions{. These terms are}
\begin{equation} \label{INIT}
	u_0(t) = \mp\i \mathrm{e}^{\i\left(\left(A^2-V^2\right)t/2 +Vx_{\pm}\right)} \quad\quad \text{and} \quad\quad v_0(t) = \mp\i\mathrm{e}^{-\i\left(\left(A^2-V^2\right)t/2 + Vx_{\pm}\right)}\,, 
\end{equation}
where the upper and lower sign choices correspond.

We substitute the expansions \eqref{INEX} into the inner {expressions} \eqref{INNEu} and \eqref{INNEv}, and we match terms of size {$\mathcal{O}\left(\eta^{-2j-3}\right)$} as $\eta\rightarrow\infty$. This gives the following system of recurrence relations:
\begin{equation} \label{REC} 
	\left(\frac{1}{2}\frac{(2j+2)!}{(2j)!}+2u_0v_0\right)u_j + u_0^2v_j = -\frac{\lambda}{2}\frac{(2j+2)!}{(2j-2)!}u_{j-1} - \sum_{k=1}^{j-1}\sum_{l=0}^{k}u_lv_{k-l}u_{j-k} - \sum_{l=1}^{j-1}u_lv_{j-l}u_0\,, 
\end{equation}
 \begin{equation}  \label{RECC}
		v_0^2u_j + \left(\frac{1}{2}\frac{(2j+2)!}{(2j)!}+2u_0v_0\right)v_j = -\frac{\lambda}{2}\frac{(2j+2)!}{(2j-2)!}v_{j-1} - \sum_{k=1}^{j-1}\sum_{l=0}^{k}v_lu_{k-l}v_{j-k} - \sum_{l=1}^{j-1}v_lu_{j-l}v_0\,.
\end{equation}
Solving equations \eqref{REC} and \eqref{RECC} yields
\begin{equation}  \label{labernew}
	u_j = (-1)^j\lambda^jc_ju_0  \quad \text{ and } \quad v_j = (-1)^j\lambda^jc_jv_0\,, 
\end{equation} 
where 
\begin{equation}
	c_j\left(\frac{1}{2} \frac{(2j+2)!}{(2j)!}-3\right) = \frac{1}{2}\frac{(2j+2)!}{(2j-2)!}c_{j-1} + \sum\limits_{k=1}^{j-1}\sum\limits_{l=0}^{k}c_lc_{k-l}c_{j-k}+\sum\limits_{l=1}^{j-1}c_lc_{j-l}\,,
\end{equation} 
with $c_0=1$ and $c_1=4$. 

\begin{figure}[tb]
	\begin{center}
		\includegraphics[width = 0.65\linewidth]{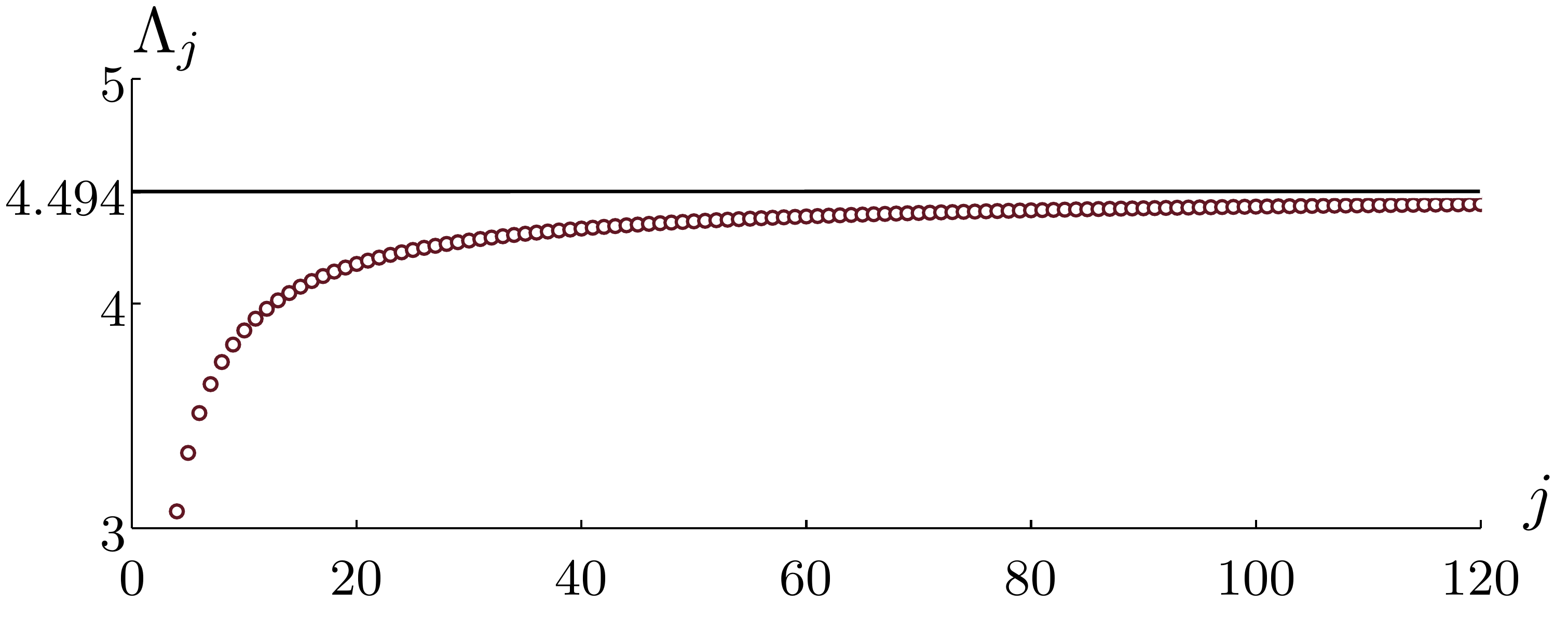} ~~~~~
	\end{center}
	\caption{Our asymptotic approximation of the limit $\Lambda=\lim_{j\rightarrow\infty}\Lambda_j$ for progressively larger values of $j$ demonstrates that the approximation converges to	$\Lambda \approx 4.494$.}
	\label{LIMIT}
\end{figure}

Motivated by comparing the prefactor in \eqref{PREFF} to \eqref{labernew} and the first term in the recurrence relation \eqref{INIT}, we set
\begin{equation}
	f(t) = \alpha \Lambda \mathrm{e}^{\i\left(\left(A^2-V^2\right)t/2 +Vx_{\pm}\right)} \, ,
\end{equation}
where we include the constant $\alpha$ for subsequent algebraic convenience and we determine the constant $\Lambda$ using asymptotic matching between the outer and inner solutions. By matching the late-order-terms \eqref{LOT} with the inner problem \eqref{labernew}, we see that $\Lambda = \lim_{j\to\infty} c_j/\Gamma(2j+1)$. We approximate $\Lambda$ by computing $c_j$ for large $j$. In Figure \ref{LIMIT}, we show the results of this approximation up to $j = 120$. We compute values of $c_j$ up to $j = 1400$; this is sufficient to obtain three decimal places of accuracy in the asymptotic approximation, which gives $\Lambda \approx 4.494$. 

We determine $f(t)$ by matching the inner solution \eqref{INEX} in the outer limit $\eta \rightarrow \infty$ with the {late-order terms} \eqref{LOT} in the inner limit $j\rightarrow\infty$ and $x\rightarrow x_{\pm}$. For the two singularities $x=x_+$ and $x=x_-$, we obtain
\begin{equation} \label{inpref}
	f_1(t) = \alpha\Lambda\a^{\i\left(\left(A^2-V^2\right)t/2 + 2Vx_{+}\right)}
	\quad \text{ and } \quad  f_2(t) = \alpha\Lambda\a^{\i\left(\left(A^2-V^2\right)t/2 + 2Vx_{-}\right)}\,. 
\end{equation}

We have now completely determined the form of the prefactors, so we have derived expression \eqref{fullLOT} for the asymptotic behavior of the late-order terms of \eqref{KARP}.

\subsection{Calculations of Remainders} \label{bapkar}


\subsubsection{Optimal Truncation} \label{eaopt}

To apply exponential asymptotic analysis, we need to determine the optimal truncation point of an asymptotic series expansion. An asymptotic series \eqref{asyexp} breaks down when consecutive terms in the series have the same size in the asymptotic limit. This typically occurs after an asymptotically large number of term~\cite{boyd1999devil}. Therefore, we determine the optimal truncation point using the late-order terms \eqref{LOT}. We find that $N_{\mathrm{opt}}\sim|\chi|/2\epsilon$ as $\epsilon \rightarrow 0$, which validates the assumption that $N_{\mathrm{opt}}\rightarrow \infty$ as $\epsilon\rightarrow 0$. Therefore, we set $N_{\mathrm{opt}} = |\chi|/2\epsilon + \omega$, where we choose $\omega\in[0,1)$ to ensure that $N_{\mathrm{opt}}$ is an integer.


\subsubsection{Stokes-Switching Analysis} \label{CRC}

Optimally truncating the asymptotic series \eqref{asyexp} gives
\begin{equation} \label{remanna}
    \psi(x,t) = \sum\limits_{j=0}^{N_{\mathrm{opt}}-1}\epsilon^{2j}\psi_j(x,t) + R(x,t)\,,
\end{equation}
where $R$ is exponentially small as $\epsilon\to 0$. We apply \eqref{remanna} to the continuous fourth-order Karpman equation \eqref{KARP} to obtain
\begin{align} \label{bigsers}
	\i \pdiff{R}{t}  + \frac{1}{2}\pdiff{^2 R}{x^2} & + \frac{\lambda\epsilon^2}{2}\pdiff{^4 R}{x^4} = -\i\sum_{j=0}^{N_{\mathrm{opt}}-1}\epsilon^{2j}\pdiff{\psi_j}{t} -\frac{1}{2}\sum_{j=0}^{N_{\mathrm{opt}}-1}\epsilon^{2j}\pdiff{^2 \psi_j}{x^2} \nonumber \\ & ~~~~~~~~~~~~~~~~~~~~~~~~~~~ -\frac{\lambda\epsilon^2}{2}\sum_{j=0}^{N_{\mathrm{opt}}-1}\epsilon^{2j}\pdiff{^4 \psi_j}{x^4} - \left|\sum_{j=0}^{N_{\mathrm{opt}}-1} \left(\epsilon^{2j}\psi_j + {R}\right)\right|^2\sum_{j=0}^{N_{\mathrm{opt}}-1} \left(\epsilon^{2j}\psi_j + {R}\right)\,. 
\end{align}
We now use the recurrence relation \eqref{epsmatch} to simplify \eqref{bigsers}. In the limit $\epsilon\rightarrow0$, we obtain the leading-order terms  
\begin{equation} \label{STOK}
	\i \pdiff{R}{t} + \frac{1}{2}\pdiff{^2 R}{x^2} + \frac{\lambda\epsilon^2}{2}\pdiff{^4 R}{x^4} \sim \frac{\epsilon^{2N_{\mathrm{opt}}}}{2}\pdiff{^2 \psi_{N_{\mathrm{opt}}}}{x^2} \,, 
\end{equation}
where the omitted terms are at most $\mathcal{O}(\epsilon^{2N_{\mathrm{opt}}+1})$ {in this limit and thus do}
not contribute to the subsequent analysis.

The optimal truncation occurs for large $N_{\mathrm{opt}}$ as $\epsilon\rightarrow0$, so we apply the late-order ansatz \eqref{LOT} to \eqref{STOK} {and obtain} 
\begin{equation} \label{REMSIM} 
	\i \pdiff{R}{t} + \frac{1}{2}\pdiff{^2 R}{x^2} + \frac{\lambda\epsilon^2}{2}\pdiff{^4 R}{x^4}  \sim \frac{\epsilon^{2N_{\mathrm{opt}}}}{2}\frac{\Psi\Gamma(2N_{\mathrm{opt}}+3)\chi_x^2}{\chi^{2N_{\mathrm{opt}}+3}} \quad \text{as} \quad \epsilon\rightarrow 0\,. 
\end{equation}
The term on the right-hand side of \eqref{REMSIM} is smaller than the terms on the left-hand side everywhere except in a neighborhood of width $\mathcal{O}(\sqrt{\epsilon})$ around {the curve $\mathrm{Im}(\chi) = 0$. This is {a} Stokes {curve}. }

We determine the behavior away from the Stokes curve by solving the homogeneous equation 
\begin{equation} \label{REMSIMHOM}
		\i \pdiff{R}{t} + \frac{1}{2}\pdiff{^2 R}{x^2} + \frac{\lambda\epsilon^2}{2}\pdiff{^4 R}{x^4}  = 0\,. 
\end{equation}
 Using a Liouville--Green (i.e., WKBJ) analysis on equation \eqref{REMSIMHOM} allows us to determine the behavior of the remainder {around the} Stokes curve using the ansatz
\begin{equation} \label{REMAN}
	{R}(x,t) \sim \mathcal{S}(x,t)\Psi(x,t)\a^{-\chi(x,t)/\epsilon}\quad \mathrm{as} \quad \epsilon\rightarrow 0\,,
\end{equation}
where $\mathcal{S}(x,t)$ is the Stokes multiplier, which varies rapidly in a region around the Stokes curve {that is associated with $\chi$, but it} tends to a different constant on the two sides of the Stokes curve. {We use $\mathcal{S}_{\nu}$ to denote
the Stokes multiplier that is associated with the singulant $\chi_{\nu}$ for $\nu \in \{1,2\}$.}

Substituting the ansatz \eqref{REMAN} into \eqref{REMSIM} gives
\begin{align} \label{REMAS}
		\frac{1}{\epsilon^{2}}\left(\frac{\lambda}{2}\chi_x^4+ \frac{1}{2}\chi_x^2\right)\Psi\mathcal{S}
		 &-  \frac{1}{\epsilon} \left(\i\Psi\chi_t+\Psi_x\chi_x+2\lambda\Psi_x \chi_x^3 \right)\mathcal{S} \nonumber \\
		  &- \frac{1}{\epsilon} \left(\chi_x + 2\lambda\chi_x^3\right)\Psi\pdiff{\mathcal{S}}{x} \sim  \frac{\epsilon^{2N_{\mathrm{opt}}}}{2}\frac{\Psi\Gamma(2N_{\mathrm{opt}}+3)\chi_x^2}{\chi^{2N_{\mathrm{opt}}+3}}\a^{\chi/\epsilon} \quad \text{as} \quad \epsilon\rightarrow 0\,. 
\end{align} 
We simplify the expression \eqref{REMAS} by using the singulant equation \eqref{SING} and prefactor equation \eqref{PRE} to obtain
\begin{equation} \label{SDX}
	\pdiff{\mathcal{S}}{x} \sim \frac{\epsilon^{2N_{\mathrm{opt}}+1}}{2}\frac{\Gamma(2N_{\mathrm{opt}}+3)\chi_x}{\chi^{2N_{\mathrm{opt}}+3}}\a^{\chi/\epsilon} \quad \text{as} \quad \epsilon\rightarrow 0\,. 
\end{equation}

It is useful to change variables to make $\chi$ the independent variable. 
We write $\mathcal{S}_x = \mathcal{S}_\chi\chi_x$, where the subscripts indicate partial differentiation. Equation \eqref{SDX} then becomes
\begin{equation} \label{SDCHI} 
	\pdiff{\mathcal{S}}{\chi} \sim \frac{\epsilon^{2N_{\mathrm{opt}}+1}}{2}\frac{\Gamma(2N_{\mathrm{opt}}+3)}{\chi^{2N_{\mathrm{opt}}+3}}\a^{\chi/\epsilon} \quad \text{as} \quad \epsilon\rightarrow 0\,. 
\end{equation}
The optimal truncation point is $N_{\mathrm{opt}} = |\chi|/2\epsilon + \omega$. Substituting the optimal truncation point into \eqref{SDCHI} shows that the right-hand side is exponentially small everywhere except in a small region around $\Arg(\chi)=0$, which corresponds to the Stokes curve. 

Following Olde Daalhuis et al.~\cite{olde1995stokes}, we define a region of width $\mathcal{O}(\sqrt{\epsilon})$ around
the Stokes curve and we obtain an inner expression for $\mathcal{S}$ in this region. Solving \eqref{SDCHI} gives 
\begin{equation} \label{CONS}
	\mathcal{S} \sim\frac{\i\pi}{2\epsilon}\erf\left(\sqrt{\frac{|\chi|}{2\epsilon}}\Arg(\chi)\right) + C \quad \text{as} \quad \epsilon\rightarrow 0\,, 
\end{equation}
 where $C$ is an arbitrary integration constant. 

Crossing the Stokes curve for $\chi_1$ from the left (i.e., when ${\Arg}(\chi_1) < 0$) to the right (i.e., when ${\Arg}(\chi_1) > 0$) causes the Stokes multiplier to jump by a finite value. We denote the change in the multiplier $\mathcal{S}_1$ by
\begin{equation} \label{JUMP0}	
	\left[\mathcal{S}_1\right]_-^+ \sim \i\pi/\epsilon \quad \text{as} \quad \epsilon \rightarrow 0\,, 
\end{equation}
where the notation $\left[f\right]_-^+$ indicates the change in the quantity $f$ from crossing the Stokes curve.
 Crossing the Stokes curve for $\chi_2$ from the left (i.e., when ${\Arg}(\chi_2) > 0$) to the right (i.e., when ${\Arg}(\chi_2) < 0$) causes the Stokes multiplier $\mathcal{S}_2$ to jump by
\begin{equation} \label{JUMP1}
	\left[\mathcal{S}_2\right]_-^+ \sim -\i\pi/\epsilon \quad \text{as} \quad \epsilon\rightarrow 0\,. 	
\end{equation}
Therefore, the changes in the remainder terms \eqref{REMAN} from crossing the Stokes curves from left to right are
\begin{align}
	\left[R_{1}\right]_-^+ &\sim \frac{\i\pi\alpha\Lambda}{\epsilon}\a^{-\alpha\pi/(2A\epsilon)}\a^{\i\left(A^2+V^2\right)t/2}\a^{-V\pi/A}\quad \text{as} \quad \epsilon\rightarrow 0 \,, \\
	\left[R_{2}\right]_-^+ &\sim -\frac{\i\pi\alpha\Lambda}{\epsilon}\a^{-\alpha\pi/(2A\epsilon)}\a^{\i\left(A^2+V^2\right)t/2}\a^{V\pi/A}\quad \text{as} \quad \epsilon\rightarrow 0 \,.
\end{align}
In Figure \ref{stokmult}, we showed the jumps in the Stokes multipliers from crossing the Stokes curves.


\section{Detailed Calculations for the CDK2 Equation}

We now give detailed calculations for obtaining the late-order terms and exponentially small contributions of the CDK2 equation \eqref{TSIN}.


\subsection{Local Analysis Near Singularities} \label{CTPP}

We determine the time-dependent behavior of the prefactor $f(t)$ \eqref{CDPRE} of the late-order terms \eqref{LOT} of the CDK2 equation \eqref{TSIN} by matching the late-order terms \eqref{LOT} to an inner problem near each singularity $x=x_{\pm}$. The late-order terms \eqref{LOT} break down in the expansion \eqref{asyexp} if $\epsilon^{2j}\psi_j \sim \epsilon^{2j+2}\psi_{j+1}$ as $\epsilon \to 0$. This occurs when $x-x_{\pm} = \mathcal{O}(\epsilon)$. We define inner variables by setting $\epsilon\eta=x-x_{\pm}$ and  $\phi(\eta,t)=\epsilon\psi(x,t)$. We retain the leading-order terms, which are of size $\mathcal{O}(\epsilon^{-3})$ as $\epsilon\rightarrow 0$. The rescaled CDK2 equation \eqref{CDNLS4} near the singularities becomes
\begin{equation} \label{CDIN}
	(1-4{\lambda})\sum_{r=0}^\infty\frac{1}{(2r+2)!}\frac{\partial^{2r+2}}{\partial\eta^{2r+2}}\phi + {\lambda}	\sum_{r=0}^\infty\frac{2^{2r+2}}{(2r+2)!}\frac{\partial^{2r+2}}{\partial\eta^{2r+2}}\phi + \phi\overline{\phi}\phi = 0\,,
\end{equation}
where the neglected terms are at most $\mathcal{O}(\epsilon^{-2})$. 

We define new variables $U(\eta,t)$ and $V(\eta,t)$ in the same fashion as in Appendix \ref{CTP}, and we obtain a coupled system of equations that is similar to \eqref{INNEu}--\eqref{INNEv}. Expanding these variables using the series in \eqref{INEX} and matching terms of size {$\mathcal{O}\left(\eta^{-2j-3}\right)$} as $\eta\rightarrow\infty$ gives a system of recurrence relations for $u_j$ and $v_j$. The solution of these recurrence relations is 
\begin{equation} \label{LIM21}
	u_j({\lambda}) = (-1)^j c_j({\lambda})u_0\,, \qquad v_j({\lambda}) = (-1)^j c_j({\lambda})v_0 \,,
\end{equation}	
where $u_0$ and $v_0$ are given in equation \eqref{INIT} and $c_j({\lambda})$ satisfies the recurrence relation 
\begin{equation}	
	c_j({\lambda}) = \frac{\sum\limits_{k=1}^{j}(-1)^k\frac{({\lambda}(4-2^{2k+2})-1)(2j+2)!}{(2k+2)!(2j-2k)!}c_{j-k}({\lambda})  + \sum\limits_{k=1}^{j-1}\sum\limits_{l=0}^{k}c_l({\lambda})c_{k-l}({\lambda})c_{j-k}({\lambda})+\sum\limits_{l=1}^{j-1}c_l({\lambda})c_{j-l}({\lambda})}{\frac{1}{2} \frac{(2j+2)!}{(2j)!}-3}\,,
\end{equation} 
with $c_0=1$ and $c_1 = (1+12{\lambda})/3$.

\begin{figure}[tb]
	\begin{center}
		\includegraphics[width = 0.65\linewidth]{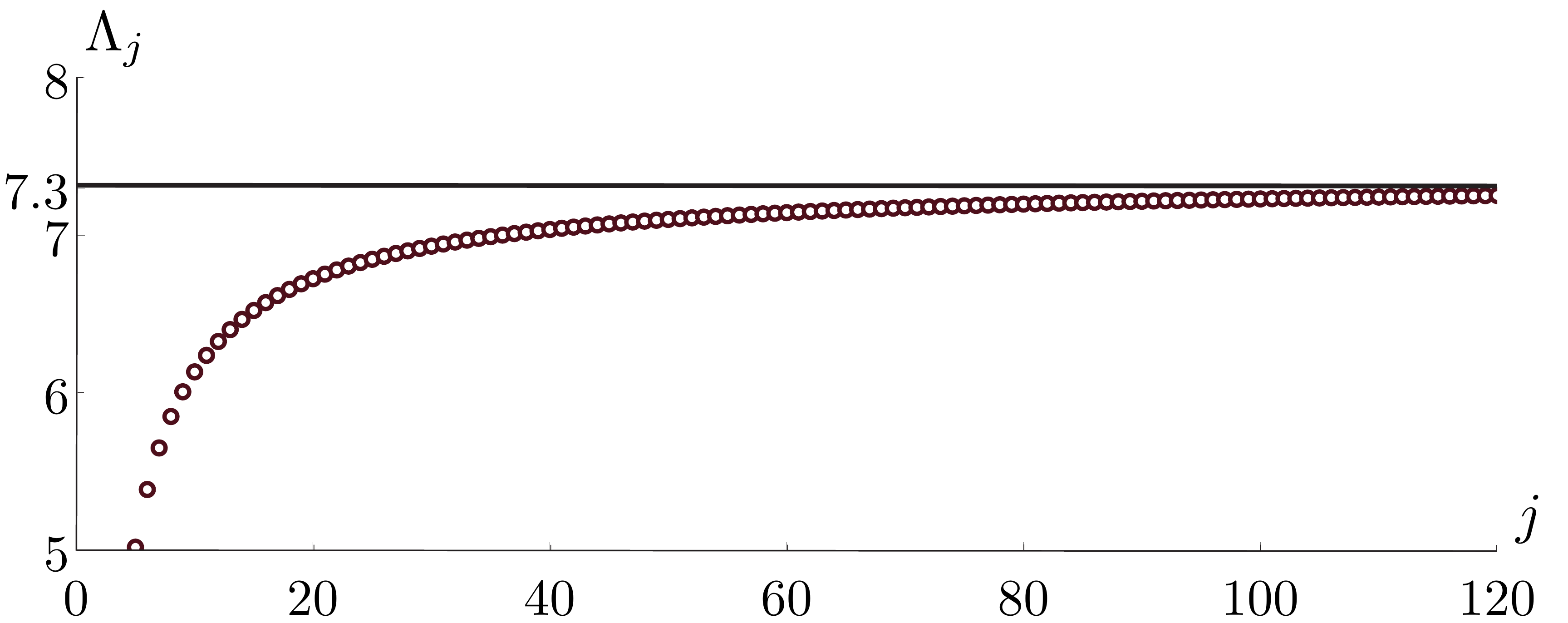} ~~~~~
	\end{center}
	\caption{Our asymptotic approximation of the limit $\Lambda=\lim_{j\rightarrow\infty}\Lambda_j(1)$ for progressively larger values of $j$ demonstrates that the approximation converges to the constant $\Lambda \approx 7.3$
	}
	\label{LIMIT2}
\end{figure}

Motivated by comparing the prefactor in \eqref{PREFF} to \eqref{cdinb} and using $u_0$ and $v_0$ from \eqref{INIT}, we set
\begin{equation}
	f(t) = \Lambda(\lambda) 
	\exp\left\{{\i\left[\left(\frac{A^2-V^2}{2}\right)t +\left(V + \frac{\sin(\sigma V\beta({\lambda}))}{\sigma \sin(\beta({\lambda}))}\right)x_{+}\right]}\right\} \,.
\end{equation}
We introduce a new variable $\Lambda_j({\lambda}) = c_j({\lambda})\beta({\lambda})^{2j+1}/\Gamma(2j+1)$. We can now numerically evaluate the limit $\Lambda({\lambda})=\lim_{j\rightarrow\infty}\Lambda_j({\lambda})$ for any given ${\lambda}$. In Figure \ref{LIMIT2}, we demonstrate our numerical evaluation of $\Lambda_j$ for ${\lambda}=1$ up to $j = 120$. {We compute values of $\Lambda_j(1)$ up to $j = 800$, which is sufficient to obtain one decimal place of accuracy in the 
asymptotic approximation, which gives 
$\Lambda\approx 7.3$.}

We determine $f(t)$ by matching the inner solution \eqref{INEX} in the outer limit $\eta \rightarrow \infty$ with the late-order terms \eqref{LOT} in the inner limit $j\rightarrow\infty$ and $x\rightarrow x_{\pm}$. We obtain
\begin{align} \label{cdinbap}
	f_1(t) = \Lambda({\lambda})
	\exp\left\{{\i\left[\left(\frac{A^2-V^2}{2}\right)t +\left(V + \frac{\sin(\sigma V\beta({\lambda}))}{\sigma \sin(\beta(\lambda))}\right)x_{+}\right]}\right\} \,, \notag \\	
	f_2(t) = \Lambda({\lambda})
	\exp\left\{{\i\left[\left(\frac{A^2-V^2}{2}\right)t +\left(V + \frac{\sin(\sigma V\beta(\lambda))}{\sigma \sin(\beta(\lambda))}\right)x_{-}\right]}\right\}\,. 
\end{align}
We now have completely determined the form of the prefactors. This yields the expression \eqref{CDfullLOT} for the asymptotic behavior of the late-order terms of \eqref{TSIN}.

\subsection{Calculations of Remainders} \label{bapkar2}

\subsubsection{Stokes-Switching Analysis}

As in Section \ref{eaopt}, we apply the optimal-truncation heuristic to identify the optimal truncation point of the series {\eqref{eaasympser}} 
as $N_{\mathrm{opt}} = |\chi|/2\epsilon + \omega$, 
where we choose $\omega\in[0,1)$ to ensure that $N_{\mathrm{opt}}$ is an integer.

Substituting the optimally truncated series {\eqref{REMso}} into the CDK2 equation \eqref{CDNLS4} yields
{\begin{align} \label{CDNLS4rem}
	&\i \sum_{r=0}^{\infty} \frac{(\sigma \epsilon)^{2r}}{(2r+1)!}\frac{\partial^{2r+1}}{\partial t^{2r+1}}\sum_{j=0}^{N_{\mathrm{opt}}-1} \left(\epsilon^{2j}\psi_j + R\right) + \left(1-4\lambda\right)\sum_{r=0}^{\infty}\frac{\epsilon^{2r}}{(2r+2)!} 
\frac{\partial^{2r+2}}{\partial x^{2r+2}}\sum_{j=0}^{N_{\mathrm{opt}}-1}\left( \epsilon^{2j}\psi_j + R\right) \nonumber \\ &~~~~~~~~~~~~+ \lambda \sum_{r=0}^{\infty} \frac{2^{2r+2}\epsilon^{2r}}{(2r+2)!} 
\frac{\partial^{2r+2}}{\partial x^{2r+2}}\sum_{j=0}^{N_{\mathrm{opt}}-1}\left( \epsilon^{2j}\psi_j + R\right)  + \left|\sum_{j=0}^{N_{\mathrm{opt}}-1}\left( \epsilon^{2j}\psi_j + R\right)\right|^2\sum_{j=0}^{N_{\mathrm{opt}}-1}\left( \epsilon^{2j}\psi_j + R\right)  = 0\,.	
\end{align}}%
Using the recurrence relation \eqref{CDREC} to simplify \eqref{CDNLS4rem} gives the leading-order equation 
\begin{align} \label{remeq2}
	\i \sum_{r=0}^{\infty} \frac{(\sigma \epsilon)^{2r}}{(2r+1)!}\frac{\partial^{2r+1}R}{\partial t^{2r+1}}  + \left(1-4\lambda\right)\sum_{r=0}^{\infty} \frac{\epsilon^{2r}}{(2r+2)!}
\frac{\partial^{2r+2}R}{\partial x^{2r+2}} 
+ \lambda \sum_{r=0}^{\infty} \frac{2^{2r+2}\epsilon^{2r}}{(2r+2)!} 
\frac{\partial^{2r+2}R}{\partial x^{2r+2}}  \sim  \frac{\epsilon^{2N}}{2}\frac{\partial^2 \psi_{N_{\mathrm{opt}}}}{\partial x^2} \quad \text{ as } \quad \epsilon \rightarrow 0\,, 
\end{align} 
where the omitted terms are at most $\mathcal{O}(\epsilon^{2N_{\mathrm{opt}}+1})$ in the limit $\epsilon\rightarrow 0$ and do not contribute to our subsequent analysis.

We substitute the late-order ansatz \eqref{LOT} into \eqref{remeq2} to give  
\begin{align} \label{remeq22}
	&\i \sum_{r=0}^{\infty} \frac{(\sigma \epsilon)^{2r}}{(2r+1)!}\frac{\partial^{2r+1}R}{\partial t^{2r+1}} + \left(1-4\lambda\right)\sum_{r=0}^{\infty} \frac{\epsilon^{2r}}{(2r+2)!}
\frac{\partial^{2r+2}R}{\partial x^{2r+2}}\nonumber \\& ~~~~~~~~~~~~~~~~~~~~~~~~~~~~~~~~~~~~~~~~~~~~~~ + \lambda \sum_{r=0}^{\infty} \frac{2^{2r+2}\epsilon^{2r}}{(2r+2)!} 
\frac{\partial^{2r+2}R}{\partial x^{2r+2}} \sim  \frac{\Psi\Gamma(2N_{\mathrm{opt}}+3)\chi_x^2}{\chi^{2N_{\mathrm{opt}}+3}} \quad \text{ as } \quad \epsilon\rightarrow 0\,. 
\end{align}
We determine the behavior away from the Stokes curve (as in Appendix \ref{CRC}) using a Liouville--Green (i.e., WKBJ) analysis of the homogeneous version of \eqref{remeq22}. This Liouville--Green analysis suggests that the ansatz \eqref{REMAN} (see Appendix \ref{CRC}) will allow us to determine the behavior near the Stokes curve. 

We use the ansatz \eqref{REMAN} and {the expression} \eqref{remeq22}, and we then collect terms in powers of $\epsilon$ to obtain
\begin{align} \label{remcalcas}  
		& \frac{1}{\epsilon^{2}}\mathcal{S}\Psi\left(\left(1-4\lambda\right)\sum_{r=0}^{\infty} \frac{\chi_x^{2r+2}}{(2r+2)!} + \lambda\sum_{r=0}^{\infty} \frac{\left(2\chi_x\right)^{2r+2}}{(2r+2)!}\right) \nonumber\\
		& ~~~ -  \frac{1}{\epsilon}\mathcal{S} \left(\frac{\i\Psi}{\sigma}\sum_{r=0}^{\infty} \frac{(\sigma \chi_t)^{2r+1}}{(2r+1)!}+\Psi_x\left[\left(1-4\lambda\right)\sum_{r=0}^\infty \frac{\chi_x^{2r+1}}{(2r+1)!} + 2\lambda \sum_{r=0}^\infty \frac{(2\chi_x)^{2r+1}}{(2r+1)!}\right]\right) 
		 \nonumber\\ & ~~~~~~~~~ - \frac{1}{\epsilon} \Psi\pdiff{\mathcal{S}}{x} \left[\left(1-4\lambda\right)\sum_{r=0}^\infty \frac{\chi_x^{2r+1}}{(2r+1)!} + 2\lambda \sum_{r=0}^\infty \frac{(2\chi_x)^{2r+1}}{(2r+1)!}\right] \sim  \frac{\epsilon^{2N_{\mathrm{opt}}}}{2}\frac{\chi_x^2\Psi\Gamma(2N_{\mathrm{opt}}+3)\a^{\chi/\epsilon}}{\chi^{2N_{\mathrm{opt}}+3}} \quad \text{as} \quad \epsilon\rightarrow 0\,, 
\end{align}
where the omitted terms are small in comparison to the retained terms. We simplify the expression \eqref{remcalcas} by using our expressions for the singulant \eqref{SING2} and prefactor \eqref{cdpreffan}. Evaluating the infinite sums in \eqref{remcalcas} gives
\begin{equation}
		    \pdiff{\mathcal{S}}{x}\sinh(\chi_x)  \sim    \frac{\epsilon^{2N_{\mathrm{opt}}+1}}{2}\frac{\chi_x^2\Gamma(2N_{\mathrm{opt}}+3)\a^{\chi/\epsilon}}{\chi^{2N_{\mathrm{opt}}+3}} \quad \text{as} \quad \epsilon\rightarrow 0\,. \label{remcalcas3}
\end{equation}
We make $\chi$ the independent variable. 
We write $\mathcal{S}_x = \mathcal{S}_\chi\chi_x$, where subscripts denote partial differentiation. Equation \eqref{remcalcas3} becomes
\begin{equation}
		    \pdiff{\mathcal{S}}{\chi}   \sim   \frac{\epsilon^{2N_{\mathrm{opt}}+1}\chi_x}{2\sinh(\chi_x)}\frac{\Gamma(2N_{\mathrm{opt}}+3)\a^{\chi/\epsilon}}{\chi^{2N_{\mathrm{opt}}+3}} \quad \text{as} \quad \epsilon\rightarrow 0\,. \label{remcalcas5}
\end{equation}
We now substitute the optimal truncation value $N_{\mathrm{opt}} = |\chi|/2\epsilon + \omega$ into \eqref{remcalcas5}. The right-hand side of \eqref{remcalcas5} is exponentially small everywhere except near the Stokes curve at ${\Arg}(\chi)=0$. We again define an inner region of width $\mathcal{O}\left(\sqrt{\epsilon}\right)$ around the Stokes curve, and we obtain an expression for $\mathcal{S}$ in this region. Solving this inner expression and using matched asymptotic expansions to connect this solution to the behavior away from the Stokes curve gives 
\begin{equation} \label{cdCONS}
	\mathcal{S} \sim \frac{\i\pi\chi_x}{2\epsilon\sinh(\chi_x)}\erf\left(\sqrt{\frac{|\chi|}{2\epsilon}}\Arg(\chi)\right) + C \quad \text{as} \quad \epsilon\rightarrow 0\,, 
\end{equation}
where $C$ is an arbitrary integration constant.

Crossing the Stokes curve for $\chi_1$ 
from the left (i.e., when ${\Arg}(\chi_1) < 0$) to the right (i.e., when ${\Arg}(\chi_1) > 0$) causes the Stokes multiplier $\mathcal{S}_1$ to jump by
\begin{equation} \label{cdJUMP0}
	\left[\mathcal{S}_1\right]_-^+ \sim \frac{\i\pi\beta}{{\epsilon\sin(\beta)}} \quad \text{as} \quad \epsilon\rightarrow 0\,. 	
\end{equation}
Crossing the Stokes curve for $\chi_2$ 
from the left (i.e., when ${\Arg}(\chi_2)>0$) to the right (i.e., when ${\Arg}(\chi_2)<0$) 
causes the Stokes multiplier $\mathcal{S}_2$ to jump by
\begin{equation} \label{cdJUMP1}
		\left[\mathcal{S}_2\right]_-^+ \sim -\frac{\i\pi\beta}{{\epsilon\sin(\beta)}}  \quad \text{as} \quad \epsilon\rightarrow 0\,. 
\end{equation}
Therefore, the changes in the remainder terms \eqref{REMAN} from crossing the Stokes curves from left to right are
\begin{align}
	\left[R_{1}\right]_-^+ &\sim \frac{\i\pi\beta\Lambda(\lambda)}{\epsilon\sin(\beta)}\a^{-\beta\pi/(2A\epsilon)}\a^{\i\left(A^2+V^2\right)t/2}\a^{-\pi(V+\sin(\sigma V \beta)/\sigma \sin(\beta))/(2A)}\quad \text{as} \quad \epsilon\rightarrow 0 \,, \\
	\left[R_{2}\right]_-^+ &\sim -\frac{\i\pi\beta\Lambda(\lambda)}{\epsilon\sin(\beta)}\a^{-\beta\pi/(2A\epsilon)}\a^{\i\left(A^2+V^2\right)t/2}\a^{\pi(V+\sin(\sigma V \beta)/\sigma \sin(\beta))/(2A)}\quad \text{as} \quad \epsilon\rightarrow 0 \,.
\end{align}
Combining these expressions for the remainders gives our expression for $\psi_\text{exp}$ in \eqref{cdEXPTERSYM}.


\bibliography{EAbib_revision03}{}

\end{document}